\documentclass[journal]{vgtc}                % final (journal style)
\ifpdf%                                % if we use pdflatex
  \pdfoutput=1\relax                   % create PDFs from pdfLaTeX
  \usepackage{graphicx}                % allow us to embed graphics files
  \DeclareGraphicsExtensions{.pdf,.png,.jpg,.jpeg} % for pdflatex we expect .pdf, .png, or .jpg files
\else%                                 % else we use pure latex
  \usepackage{graphicx}                % allow us to embed graphics files
  \DeclareGraphicsExtensions{.eps}     % for pure latex we expect eps files
\fi%

\graphicspath{{figures/}{pictures/}{images/}{./}} % where to search for the images

\usepackage{microtype}                 % use micro-typography (slightly more compact, better to read)
\PassOptionsToPackage{warn}{textcomp}  % to address font issues with \textrightarrow
\usepackage{textcomp}                  % use better special symbols
\usepackage{mathptmx}                  % use matching math font
\usepackage{times}                     % we use Times as the main font
\usepackage{cite}                      % needed to automatically sort the references
\usepackage{tabu}                      % only used for the table example
\usepackage{booktabs}                  % only used for the table example
\usepackage{enumitem}
\usepackage{CJK}
\usepackage{algorithm}
\usepackage{algorithmicx}
\usepackage{algpseudocode}          
\usepackage{tabularx, ragged2e}
\usepackage{color}
\usepackage{amsmath}
\floatname{algorithm}{Algorithm}

\usepackage{makecell}
\usepackage{multirow}
\usepackage{diagbox}

\onlineid{1274}

\vgtccategory{Research}
\vgtcpapertype{Algorithm/Technique}

\title{Topology Density Map for Urban Data Visualization and Analysis}

\author{Zezheng Feng, Haotian Li, Wei Zeng, Shuang-Hua Yang, and Huamin Qu}
\authorfooter{
\item
 Z. Feng H. Li and H. Qu are with the Hong Kong University of Science and Technology. E-mail: \{zfengak, haotian.li\}@connect.ust.hk and huamin@cse.ust.hk
\item
 W. Zeng is with Shenzhen Institutes of Advanced Technology, Chinese Academy of Sciences. Wei Zeng is the corresponding author. E-mail: wei.zeng@siat.ac.cn. 
\item
 SH. Yang is with Southern University of Science and Technology. E-mail: yangsh@sustech.edu.cn.
}

\abstract{
Density map is an effective visualization technique 
for depicting the scalar field distribution in 2D space.
Conventional methods for constructing density maps are mainly based on Euclidean distance, limiting their applicability in urban analysis that shall consider road network and urban traffic.
In this work, we propose a new method named \emph{Topology Density Map}, targeting for accurate and intuitive density maps in the context of urban environment.
Based on the various constraints of road connections and traffic conditions, the method first constructs a directed acyclic graph (DAG) that propagates nonlinear scalar fields along 1D road networks.
Next, the method extends the scalar fields to a 2D space by identifying key intersecting points in the DAG and calculating the scalar fields for every point, yielding a weighted Voronoi diagram like effect of space division.
Two case studies demonstrate that the Topology Density Map supplies accurate information to users and provides an intuitive visualization for decision making. An interview with domain experts demonstrates the feasibility, usability, and effectiveness of our method. 
} % end of abstract


\CCScatlist{ % not used in journal version
 \CCScat{K.6.1}{Management of Computing and Information Systems}%
{Project and People Management}{Life Cycle};
 \CCScat{K.7.m}{The Computing Profession}{Miscellaneous}{Ethics}
}

\teaser{
  \centering
  \includegraphics[width=\linewidth]{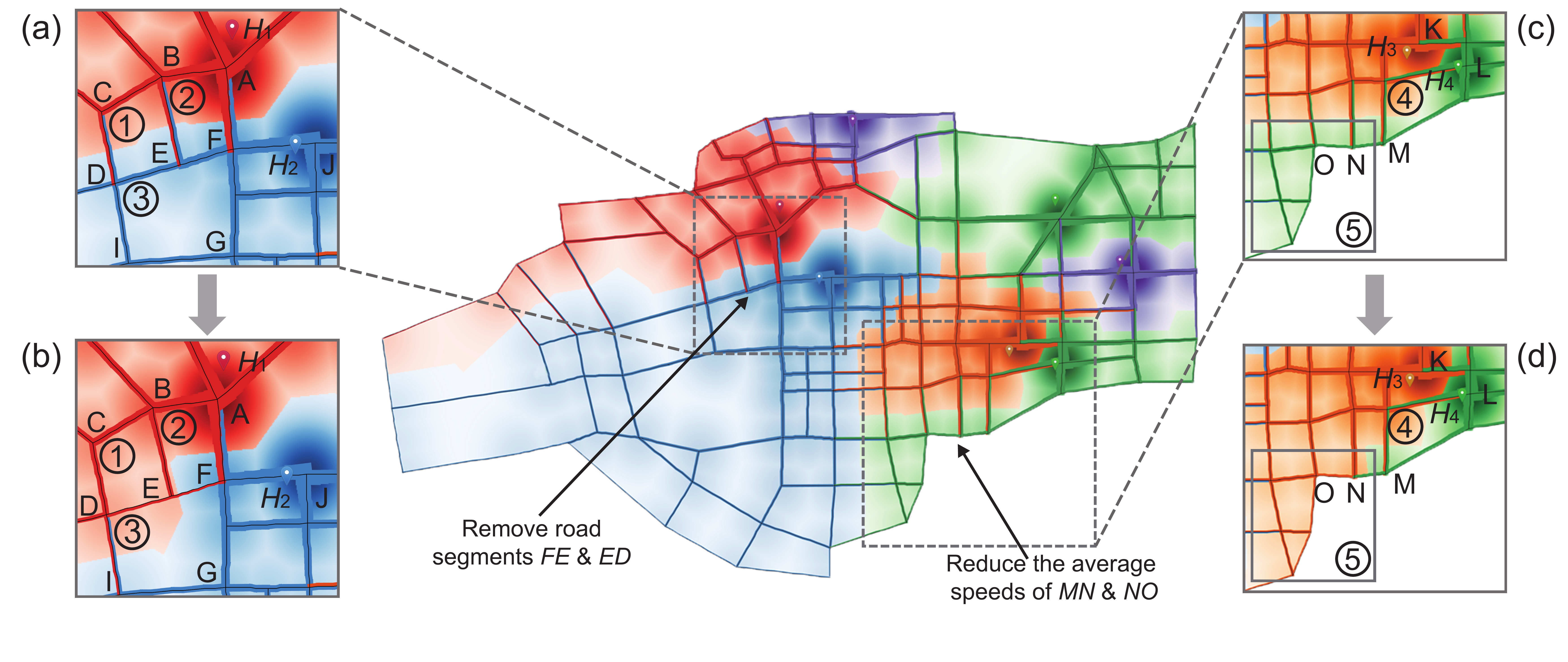}
  \vspace{-3em}
  \caption{Analyzing the influence of road topology and traffic conditions on POI accessibility: density distribution before (a) and after (b) removing the selected road segments $FE$ and $ED$; and density distribution before (c) and after (d) slowing down the average speeds of the selected road segments $MN$ and $NO$.
  }
  \label{fig:teaser}
}

\vgtcinsertpkg

\newcommand{\zw}[1]{{\color{black}{#1}}}
\newcommand{\fzz}[1]{{\color{black}{#1}}}

\begin{document}


\firstsection{Introduction}

\maketitle

%Road is a kind of infrastructure connecting different areas for both pedestrians and vehicles. In recent decades, with the expansion of the city and the evolution of transportation technology, the city's road network becomes more and more intensive and unobstructed, so the two geographically far away areas may have short commuting time due to the road connectivity. At the same time, it also makes the impact of many public service facilities on the general public radiate along with the complex road network\cite{jin2019evaluating} to the areas which are geographically far away. Moreover, the accessibility plays an important role in the area of urban planning\cite{de2013cities}.%

Road network is an important infrastructure in cities.%~\cite{djahel2019wave}. 
~With expansion of cities and evolution of transportation technology, road networks are becoming increasingly intensive and unobstructed.
Consequently, two geographically distant locations may have a short commuting time due to road connectivity~\cite{baddeley2020analysing}.
Hence, road network is of importance in many real-world applications, including urban planning~\cite{talebi2019gis} and \zw{urban scaling analysis~\cite{li2017simple}}.
For example, traffic management can benefit from forecasting and visualizing the congestion conditions of a city based on vehicle detector data on roads~\cite{lee2019visual}.
Moreover, road network also shields an effect on urban data visualization.
For an example, a recent work of~\cite{zeng2019route} presented a more accurate edge bundling method for urban traffic data, by carefully considering the constraints implied by underlying road
networks.

Density map, as an effective visualization technique~\cite{he2019diverse}, is widely used in urban analysis, which involves spatial distribution patterns~\cite{zeng2017visualizing}, such as anomaly detection~\cite{thom2012spatiotemporal}, risk analysis~\cite{scheepens2011composite} and air pollution propagation analysis~\cite{deng2019airvis}. %, and crowd estimation~\cite{sindagi2017generating}.
A density map depicts the continuous distribution of scalar field in a 2D planar space by assigning a unique color to each individual scalar value~\cite{silverman2018density,10.1111:cgf.14031}.
The scalar field is computed from the premise of the finite observation of the data~\cite{silverman2018density}.
The process is referred as density estimation, which can be parametric or nonparametric.
Kernel density estimation (KDE) is a common nonparametric model, which typically applies a kernel (\emph{e.g.}, parabolic, Gaussian, and Sigmoid) to the proximity between two locations.
The proximity is typically computed as Euclidean distance in a Cartesian coordinate system. Due to its simplicity, KDE has been widely adopted in movement visualization, such as vessel movements~\cite{wilems_2009_visualization, scheepens2011composite} and flight trails~\cite{hurter_2009_fromdady, hurter_2014_bundled}.

% The distribution of scalar field in space is mostly discrete.
% In most cases, it is difficult for us to obtain the scalar field distribution of each location in the research region, so we have to estimate the scalar field based on the known value, which is density estimation. Density estimation, containing parametric density estimation and nonparametric density estimation, models the probability distribution of random variable on the premise of finite observation of the data~\cite{silverman2018density}.  

% First, distance topology-based, crossings, connectiveties, obstracles, etc.
% Second, directional, 
% Thrid, temporal variant,

% In recent years, a group of scholars choose a density map to analyze the hotspot analysis problem in urban area by using urban traffic data, especially taxi GPS data~\cite{zhao2019network, tang2016network, deng2019density}.
Nevertheless, the KDE based on Euclidean distance is inappropriate for many urban analyses that should take road networks into consideration, simply because most movements in cities are constrained by road networks~\cite{chen2015survey}.
% \zw{add a sentence to describe figure 1.}
\fzz{Fig.~\ref{fig:teaser} illustrates a real-world scenario, where domain experts would like to analyze how easy to access points-of-interest (POIs), \emph{i.e.}, the POI accessibility.}
For accurate accessibility measurement, we shall consider the following properties of a road network when measuring the proximity between two locations:
% \vspace{-1.5mm}
\begin{itemize}
% \vspace{-1.5mm}
\item
\emph{Topological}: Locations are connected through roads and intersections, which can be obstructed by obstacles, such as accidents or road repairs.
\fzz{For instance, if the path~($F$-$E$-$D$) in Fig.~\ref{fig:teaser}(a) is under repair, the shortest paths from $H_2$ to locations in region 1 will be altered, affecting POI accessibility in the region.
}

\vspace{-1.5mm}
\item
\emph{Directional}: Unlike the Euclidean distance, the proximity of A $\rightarrow$ B could be very different from that of B $\rightarrow$ A.
\fzz{
% According to the standard traffic rules \& regulations, a vehicle drives along a fixed side. 
In most cities, a road has two different directions, which can be modeled as two directional edges.
For example, when the aforementioned case in Fig.~\ref{fig:teaser}(a) occurs, the shortest path ($E$ $\rightarrow$ $D$) changes but that in the other direction ($D$ $\rightarrow$ $E$) remains the same.}

\vspace{-1.5mm}
\item
\emph{Temporal variant}: The proximity here is typically measured by the access time rather than the physical distance, which depends much on traffic conditions that vary over time.
\fzz{For instance, the POI accessibility in Region 5 is greatly affected by traffic conditions, \emph{e.g.}, the non-peak hours in Fig.~\ref{fig:teaser}(c) and the peak hours in Fig.~\ref{fig:teaser}(d).}

\end{itemize}

The transportation and geography communities have strived to include road networks in KDE computation, e.g.,~\cite{delso2018new, deng2019density}.
However, proximity measurement only applies to one-dimensional road networks~\cite{tang2016network}, whilst density map requires scalar field distribution in a 2D planar surface.
This limitation yields unintuitive visualizations compared with density map. 
A new method is required to derive scalar fields outside the road network.
\zw{We consider the following requirements for such a method:
i) \emph{correctness}, which should accurately coordinate with the complex road topology and the dynamic traffic conditions; and
ii) \emph{intuitiveness}, which should intuitively depict density fields on a 2D planar surface instead of a 1D road network.}

% If applying the conventional method to calculate the density distribution of the spatial plane, due to the constraints of the road network, the density distribution cannot be calculated relatively accurately.
% A more accurate and more intuitive visualization technique is needed to visualize such features. 

% \zw{describe how topology density map is constructed here}
In this study, we propose \emph{Topology Density Map} to fulfill the requirements.
% To calculate the scalar field,
The method works as follows:
First, we construct a directed acyclic graph (DAG) from each POI to all road intersections, taking road topology and traffic conditions into consideration.
Next, we find shortest paths to intersections with minimum access times, and measure the density propagation along the shortest paths.
Lastly, we extend density estimation from each intersection to the neighboring regions, and then generate a density map based on the estimated 2D scalar fields.
% In this study, we focus on how to draw a density map for urban planners on solving hotspot analysis problem. The proposed method not only take the road topology into account but also provide urban planners with a more intuitive visualization to help them perceive the density distributions of scalar field in the whole region as well. 
The main contributions of this work are listed as follows:
\begin{itemize}
\vspace{-1.5mm}
\item The topology density map extends the existing network-constrained density estimation method from a 1D road network to a 2D planar surface,
\zw{providing more correct and intuitive density field estimation in an urban area}.
% ~\footnote{\textcolor{blue}{In this paper, when we look at the road network from a macro perspective in order to simplify the representation, we define the road network in the real world as 1D road which not take the roadway width and multi-lane properties into consideration. Meanwhile, corresponding to this situation, we define the planar surface surrounded by the 1D road as 2D planar surface.}}.
% This reveals the impact of POIs in the 2D urban area.
% Therefore, the users can explore the impact accurately and intuitively. 
% To our best knowledge, it is a novel method, extending the estimation of the density from the 1D road to the 2D planar surface.

\vspace{-1.5mm}
\item The topology density map forms a weighted Voronoi diagram like effect that reflects the space partition by the POI accessibility.
\zw{This kind of visualization is especially useful when multiple POIs affect the same region.
% For multiple POIs affect the same region, this shows the range of impact from different POIs and reveals the direction of the road.
}
% \zw{hard to understand. come back later.}

\vspace{-1.5mm}
\item The feasibility and usability of the topology density map are demonstrated by two real-world case studies, and the effectiveness of the method is confirmed by expert feedbacks.
\end{itemize}

The remaining of this paper is structured as follows:
Sec.~\ref{sec:related} discuses related works in visual analytics and visualizations for urban data analysis.
Sec.~\ref{sec:overview} summarizes the limitations of existing density estimation methods and provides an overview of our approach.
Sec.~\ref{sec:proc} describes the domain problems and details of preparing input data.
Sec.~\ref{sec:method} presents the details of the topology density map construction, and the comparison of the visual effect with network KDE.
Sec.~\ref{sec:case_study} demonstrates two case studies conducted on a real-world dataset and the feedbacks from three independent domain experts.
Sec.~\ref{sec:discuss} describes the various considerations of our approach, including the parameter selection, the computing efficiency, and alternative designs, and discusses the limitations and future works.
Finally, Sec.~\ref{sec:con} concludes this paper. 
\begin{figure}[h]
  \centering
  \includegraphics[width=0.8\linewidth]{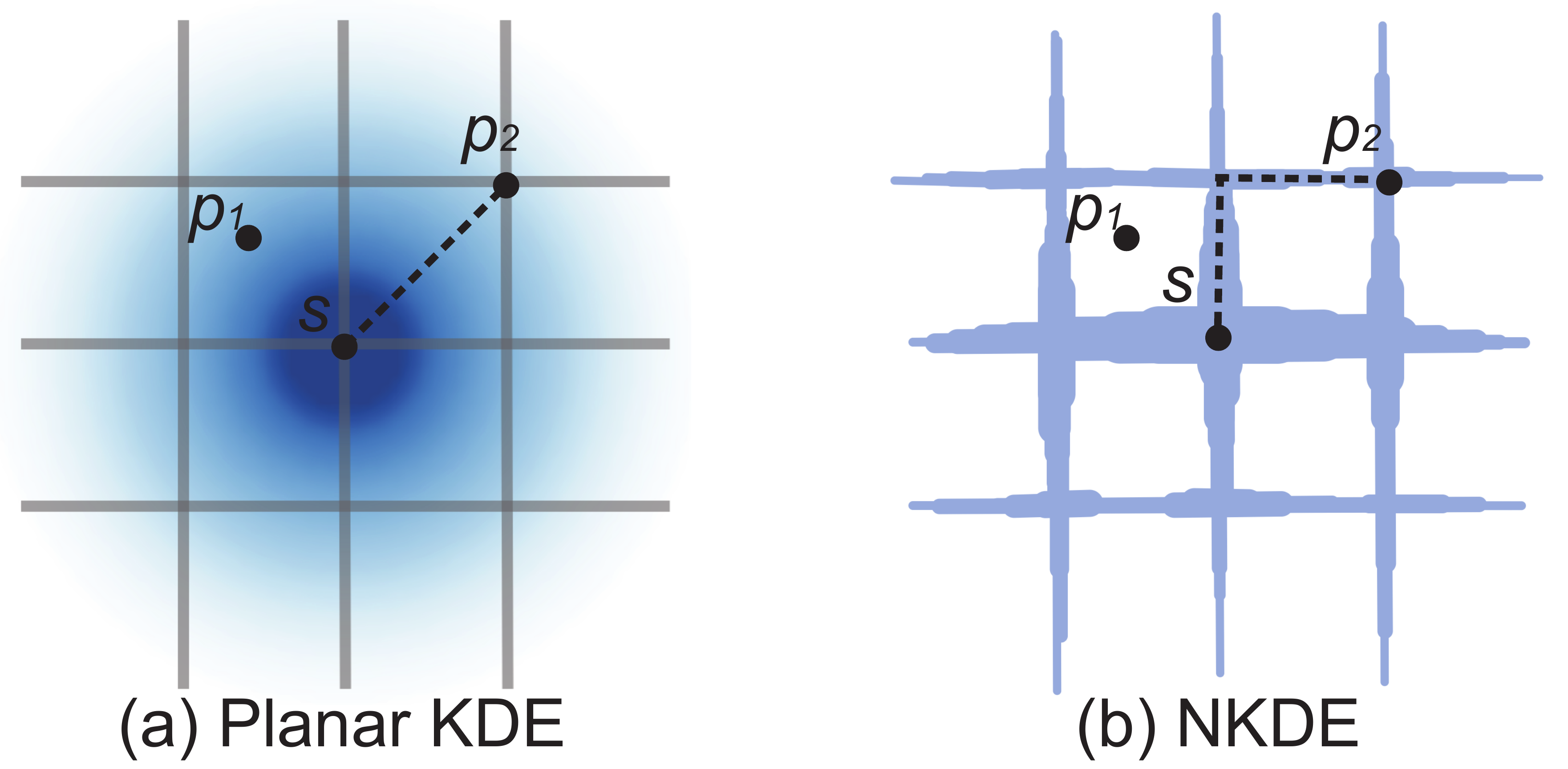}
  \vspace{-1mm}
  \caption{Illustration of density estimation using (a) planar KDE (Euclidean distance), and (b) NKDE (network distance).
  Planar KDE encodes density fields using colors, while NKDE uses line width.
  }~\label{fig:planar kde and NKDE}
  \vspace{-1em}
\end{figure}

%%%%%%%%%%%%%%%%%%%%%%%%%%%%%%%%%%%%%%%%%%%%%%%%%%%%%%
%%%%%%%%%%%%%%%%%%%%%%%%%%%%%%%%%%%%%%%%%%%%%%%%%%%%%%
\section{Related work}
\label{sec:related}

%%%%%%%%%%%%%%%%%%%%%%%%%%%%%%%%%%%%%%%%%%%%%%%%%%%%%%
%%%%%%%%%%%%%%%%%%%%%%%%%%%%%%%%%%%%%%%%%%%%%%%%%%%%%%
\subsection{Visual Analytics for Urban Traffic}

Benefiting from the advancement of location sensing technologies, movement data, which characterize moving objects in the location space (where), time space (when), and attribute space (what)~\cite{cartography}, become ubiquitous and abundant~\cite{sun2016embedding}. 
% Movement data characterize the moving objects in location space (where), time space (time), and attribute space (what)~\cite{cartography}.
Urban traffic is a specific movement data that record commuter/goods/vehicle movements on road networks in a city.
Urban traffic data provide unprecedented information on human activities~\cite{andrienko2017visual}, and visual analytics is an effective tool for analyzing urban traffic~\cite{chen2015survey}.
Many visual analytics have been developed for urban traffic in different cities for different purposes, such as to study hidden themes of taxi movements in Shenzhen~\cite{chu2014visualizing}, to make an improvement in the performance for bus route planning in Beijing~\cite{weng2020towards}, to stack and visualize trajectory attributes for movements in San Francisco~\cite{tominski2012stacking}, and to explore commuting movements in New York City \cite{ferreira2013visual}.
Interested readers may refer to the surveys~\cite{chen2015survey, andrienko2017visual}.

Visual clutter is a main issue in visual analytics for urban traffic, and it is mainly caused by large amounts of traffic data.
Many solutions have been proposed to mitigate this problem, such as designing effective filter interactions~\cite{kruger2018visual}, building efficient query database~\cite{ferreira2013visual}, and developing realistic edge bundling~\cite{zeng2019route}.
Among the solutions, density map is an effective visualization technique for depicting scalar fields on a 2D planar surface~\cite{wilems_2009_visualization, scheepens2011composite}.
Many visual analytics adopt a density map to present information in an urban environment.
For example, Shen et al.~\cite{shen2015analysis} identified hotspots of taxi movements from taxi GPS traces, which can help drivers find optimal paths between a pair of start-end locations.
Liu et al.~\cite{liu2016smartadp} revealed the space visibility from taxi GPS trajectory data, which can assist planners in identifying optimal locations for placing advertisement boards.

Both works adopted density-based epsilon distance algorithms to generate density maps.
However, the algorithms can only model the scalar field propagation at local scales, whilst many city-scale urban analyses require considerations of road network and traffic conditions, such as mobility~\cite{zeng2014visualizing} and \zw{transportation efficiency analysis~\cite{dong_2016_population}}. %accessibility~\cite{jin2019evaluating} analysis.
This work aims to fill the gap with a feasible solution for city-scale urban analysis.
We develop a new approach to density map construction, namely, \emph{Topology Density Map}.
% In our paper, we focus on the location and time stamp of the vehicles which after matching with the map, we can compute the average speed of each road.
\zw{We show the efficiency of the topology density map in city-scale accessibility analysis.}

\subsection{Density Estimation in Urban Analysis}

Scalar fields distributed over space can be visualized as a field schematization that aggregates representatives for local regions based on field data, \emph{i.e.}, density maps~\cite{10.1111:cgf.14031}.
As an effective visualization technique, density maps have been widely adopted in urban analysis~\cite{zhao2019network}.
A key step in density map construction is measuring the scalar field distribution in the study area.
Many density estimation methods have been proposed.
Among them, kernel density estimation (KDE) has been proven efficient and effective~\cite{botev2010kernel}.
On the basis of proximity measurement, the KDE for urban analysis can be divided in two parts: the density estimation based on the Euclidean distance $-$ planar KDE (Fig.~\ref{fig:planar kde and NKDE}(a)), and the density estimation based on network proximity $-$ network KDE (NKDE) (Fig.~\ref{fig:planar kde and NKDE}(b)).

Planar KDE obeys the rule that  in a Cartesion coordinate system, straight line is the shortest between two points, \emph{i.e.}, Euclidean distance.
Planar KDE is frequently employed when the trajectory position is unconstrained, such as vessel movements~\cite{wilems_2009_visualization, scheepens2011composite} and flight trails~\cite{hurter_2009_fromdady, hurter_2014_bundled}.
In cities, however, the method is only suitable for local-scale urban analyses.
For example, Grubesic~\cite{grubesic2006application} adopted planar KDE in detecting hotspots for crime in a city, and Kloog et al.~\cite{kloog2009using} applied planar KDE in studying the relationship between breast cancer cases and artificial lighting at night.
% Some researchers argue that spherical distance, instead of Euclidean distance, should be used for measuring distance between two locations in a city, because the Earth is a sphere~\cite{chang2010context}. \zw{not really useful}
For city-scale analyses that involve urban traffic along road networks, planar KDE is no longer suitable.
For an example, Zeng et al.~\cite{zeng2014visualizing} studied passenger mobility using the public transportation system in a city, which cannot be depicted using planar KDE.

The NKDE that estimates density based on the proximity along a road network is appropriate for city-scale urban analyses.
However, NKDE construction is rather challenging.
First, we need to find a suitable proximity measurement on the road network, which could be affected by many factors, such as traffic conditions, street crossings, and obstacles, etc.
For instance, Delso et al.~\cite{delso2018new} considered the impact of street crossings when measuring travel times between positions. % on a road network.
% Furthermore, Xia et al.\cite{xia2019identify} proposed a network-based spatiotemporal field clustering approach to detect the hotspot by using taxi pick and drop off data.
% They used isoline to distinguish the high-intensity area.
Second, we need to develop an efficient algorithm for computing pairwise proximities.
For example, Deng et al.~\cite{deng2019density} constructed a network-constrained Delaunay triangulation to facilitate the measurement of network proximities between locations.
Nevertheless, even when these two challenges are well resolved, density fields are computed as propagation along 1D road networks (Fig.~\ref{fig:planar kde and NKDE}(b)), which is less intuitive in comparison with density fields on a 2D planar space (Fig.~\ref{fig:planar kde and NKDE}(a)).

We develop topology density map that combines the \emph{intuitiveness} of planar KDE in presenting continuous density fields over space and the \emph{correctness} of NKDE in considering the road topology and traffic conditions for city-scale urban analysis.
Various optimization techniques are designed to facilitate the construction and visual effect of the topology density map.
\section{Requirement Analysis and Method Overview}
\label{sec:overview}

Density estimation is the key step in density map generation.
This section briefly introduces planar KDE (Sec.~\ref{sssec:planar_kde}) and NKDE (Sec.~\ref{sssec:nkde}), followed by a discussion of the requirements when applied to urban data analysis and visualization (Sec.~\ref{ssec:requirement}).
An overview of our topology density map is presented at the end (Sec.~\ref{ssec:overview_tde}).

%%%%%%%%%%%%%%%%%%%%%%%%%%%%%%%%%%%%%%%%%%%%%%%%
%%%%%%%%%%%%%%%%%%%%%%%%%%%%%%%%%%%%%%%%%%%%%%%%
\subsection{Conventional Density Estimation}

In this work, we regard planar KDE as a density estimation method that adopts kernel functions with Euclidean distance, while NKDE is a special case of planar KDE, with the distance measured on the road network as described in~\cite{xie2008kernel}.
Both approaches are state-of-the-arts methods for estimating density fields.

\subsubsection{Planar KDE}
\label{sssec:planar_kde}

Planar KDE is a nonparametric estimation method, \emph{i.e.}, not making use of prior knowledge about the data distribution. 
Instead, the method studies the data distribution characteristics from the data themselves.
Planar KDE simulates a probability distribution curve using a smooth kernel function to fit the observed data points.
The method is often used to estimate unknown probabilities, \emph{e.g.}, movement distributions in a 2D space~\cite{wilems_2009_visualization, scheepens2011composite, hurter_2009_fromdady, hurter_2014_bundled}.

Given a set of observed data points $D := \{x_i\}$, we can estimate density field $\lambda(s \, |\, D)$ at location $s$ in a 2D space as follows: 

\begin{equation}
\lambda(s \, | \, D) = \sum\limits_{x_i \in D}\frac{1}{\pi r^2}K(\frac{d(x_i - s)}{r})
\label{equ:planar_KDE}
\end{equation}
\\
where $r$ is the kernel radius, which is often called bandwidth, 
$d(\cdot)$ is the Euclidean distance between point $x_i$ and location $s$, 
and $K$ is a weighting function known as kernel function.
Several common kernel functions exist, including parabolic, Gaussian, Sigmoid, and negative exponential functions.
The function value typically decreases as the distance increases, coping with the distance-decay effect.
Given this property, planar KDE is often employed in spatial analysis, as many geographic phenomenon follow the first law of geography, which states that ``everything is related to everything else, but near things are more related than distant things"~\cite{tobler_1970_computer}.

\subsubsection{NKDE}
\label{sssec:nkde}

NKDE is usually used to estimate the density fields of road-constrained events on a road network, such as the impacts of traffic accidents.
Density fields are measured as linear propagations along a road network.
Given observed data points $D := \{x_i\}$ and road network $G$, density field $\lambda(s | D, G)$ at location $s$ can be estimated as:

\begin{equation}
\lambda(s | D, G) = \sum\limits_{x_i \in D}\frac{1}{r}K(\frac{d_{G}(x_i - s)}{r})
\label{equ:nkde}
\end{equation}
\\
where parameters $r$ and $K$ represent the same settings with those in planar KDE, while $d_{G}(\cdot)$ denotes the network proximity between $x_i$ and $s$.
In this sense, the function is only valid for location $s$ on the road network $G$.
According to Fig.~\ref{fig:planar kde and NKDE}(b), we can only estimate the density field outside the road network at $p_2$, but not for $p_1$.

\begin{figure*}[h]
\centering
  \includegraphics[width=0.95\linewidth]{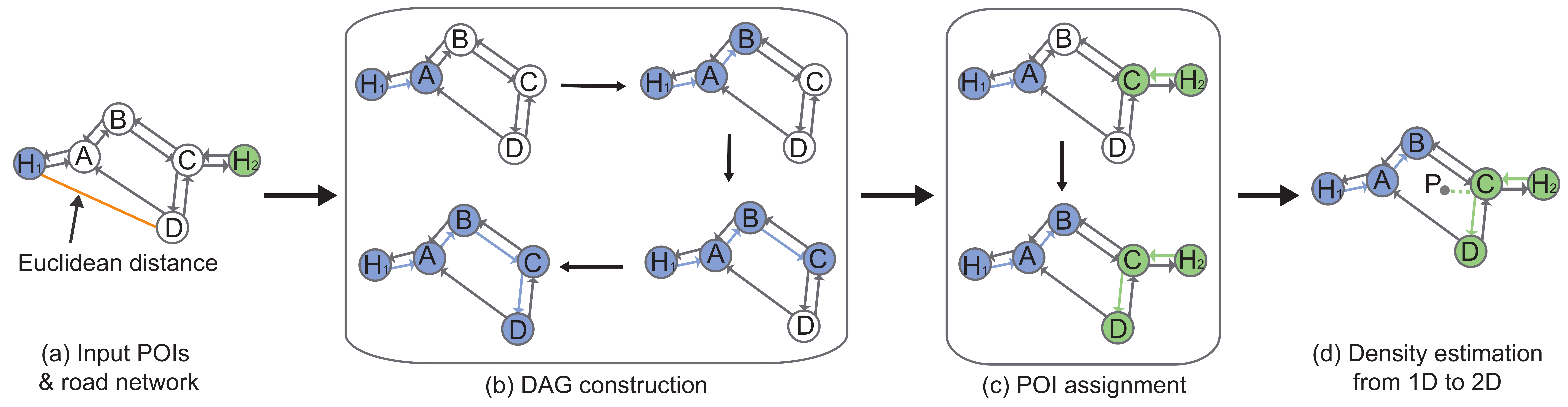}
  \vspace{-1em}
  \caption{
  Overview of topology density map construction.
  (a) Given a road network (with vertices $A$, $B$, $C$, and $D$) and multiple POIs ($H_1$, and $H_2$), we first (b) compute a directed acyclic graph (DAG) for each POI taking road topology and traffic conditions into account.
  (c) Next, we assign each vertex to a POI with the highest density field. Here $A$ \& $B$ are assigned to $H_1$, while $C$ \& $D$ are assigned to $H_2$.
  (d) Finally, we extend density estimation from 1D road network to 2D planar surface.
  % \textcolor{blue}{The differences between traditional method and Topology Density Map: the traditional method (a) estimates the density based on the Euclidean distance between two points, while there are three differences between topology density map and traditional method. Topology density map first (b) takes the road topology and the road direction into account, and then (c) estimates the density of each vertex of the POI with the shortest network distance on the road, finally (d) it extends the density estimation from 1D road to 2D planar surface.
  % }
  % \zw{is it better to call this the workflow?} 
}
  \label{fig:uniqueness}
  % \vspace{-1em}
\end{figure*}

%%%%%%%%%%%%%%%%%%%%%%%%%%%%%%%%%%%%%%%%%%%%%%%%
%%%%%%%%%%%%%%%%%%%%%%%%%%%%%%%%%%%%%%%%%%%%%%%%
\subsection{Requirement Analysis}
\label{ssec:requirement}
Though popular, planar KDE and NKDE fail to fulfill the following requirements when applied to urban data analysis.

\vspace{1.5mm}
\textbf{R1. Correctness.} 
As discussed above, planar KDE can provide an intuitive representation of the density fields over a 2D space.
The method employs Euclidean distance to estimate the density fields within the kernel radius, which however is unsuitable for urban areas.
This is because road networks in a city typically exhibit a complex topology and dynamic traffic conditions that should be considered.
There may not be a straight connection between two locations, or a straight connection may not be the fastest path due to traffic jams or road maintenance.
In such scenarios, people no longer commute via a straight line, but alternatively choose a suitable path according to road connections and current traffic conditions, such as the path between $p_2$ and $s$ in Fig.~\ref{fig:planar kde and NKDE}(b).
Furthermore, in practice, roads in a city are often \emph{bidirectional}, with median dividers separating roads into opposite directions.
On these roads, vehicles cannot cross the road or change directions until they reach turning points.
However, planar KDE cannot reflect the bidirectional characteristic when estimating density fields.
Therefore, when analyzing urban problems that consider human mobility on road networks, planar KDE will produce an inaccurate density field estimation, generating incorrect visualization that may be detrimental to decision-making. 

\vspace{1.5mm}
\textbf{R2. Intuitiveness.} 
NKDE adopts the proximity along road network instead of the Euclidean distance for estimating the density fields, which can correctly compute the impact of events (\emph{e.g.}, traffic accidents and taxi pick-ups/drop-offs) distributed along the road network.
% The density map generated by this method measures linear unit \zw{don't get the point}. 
% Hence, the method can correctly compute the impact of events (\emph{e.g.}, traffic accidents, \fzz{taxi pick-up/drop-off passengers} \zw{what is taxi behavior?}) distributed along road network.
However, this kind of density estimation can only measure the density propagation along a road network that provides 1D measurements~\cite{xie2008kernel}, unlike the density fields over a 2D space.
% draw the density map along the road which can be called a 1D linear space.
However, most human activities take place at locations not covered by road networks in a city.
NKDE fails to provide an intuitive visualization for density fields at these locations.
Taking $p_1$ in Fig.~\ref{fig:planar kde and NKDE}(b) as an example, users cannot retrieve its density field with NKDE. 
The lack of density fields hinders certain analytical tasks, such as the comparison of the density fields at $p_1$ and at $p_2$.
% show information in urban areas surrounding roads.
% Furthermore, density maps in the 1D linear space are not intuitive for people to aware the density distribution in the overall 2D geographic space.
Therefore, NKDE can only present the density fields along the road network, whilst users typically need to visualize density fields over an entire urban area.
The visualization is unintuitive, making urban analysis less effective.

\vspace{1.5mm}
\noindent
In summary, there is an urgent need for a new density estimation method that can simultaneously fulfill the requirements of \emph{correctness} and \emph{intuitiveness} in urban data visualization and analysis.

%%%%%%%%%%%%%%%%%%%%%%%%%%%%%%%%%%%%%%%%%%%%%%%%
%%%%%%%%%%%%%%%%%%%%%%%%%%%%%%%%%%%%%%%%%%%%%%%%
\subsection{Overview of Topology Density Map}
\label{ssec:overview_tde}

We develop the topology density map, a new density estimation method specifically designed for urban data visualization and analysis.
Topology density map fulfills the \emph{correctness} requirement by incorporating road topology and traffic condition into account, and the \emph{intuitiveness} requirement by extending the density estimation from a 1D road network to a 2D planar surface.

Fig.~\ref{fig:uniqueness} illustrates the construction process of the topology density map.
The method takes a road network $G$ with vertices $A$, $B$, $C$, and $D$, and two POIs $H_1$ and $H_2$ as inputs (Fig.~\ref{fig:uniqueness}(a)). 
Thanks to the open data campaign, these urban data are becoming increasingly ubiquitous, such as the open street map (OSM) for road networks.
With the information, we can construct a directed acyclic graph (DAG) for each POI, with the POI as the starting vertex and all intersections of the road network connected (Fig.~\ref{fig:uniqueness}(b)).
Here, the edges in the road network are bidirectionally modeled, and the connectivity between neighboring vertices is affected by traffic conditions.
Thus, the proximity computation is consistent with the reality.
For instance, the proximity from POI $H_1$ to vertex $D$ is computed as an aggregation of $d_{G}(HA)$, $d_{G}(AB)$, $d_{G}(BC)$, and $d_{G}(CD)$ rather than Euclidean distance between $H$ and $D$ (the yellow line in Fig.~\ref{fig:uniqueness}(a)).
Next, we assign each intersection to a POI with the highest density field (Fig.~\ref{fig:uniqueness}(c)).
Here $A$ and $B$ are assigned to $H_1$, while $C$ and $D$ are assigned to $H_2$.
Finally, we extend the density estimation from a 1D road network to a 2D planar surface (Fig.~\ref{fig:uniqueness}(d)).
For point $P$ in the region, we retrieve the shortest path to each POI by iterating on the neighboring intersections.
For instance, the proximity from $H_1$ to $P$ is $d_{G}(HA) + d_{G}(AP)$, while that from $H_2$ to $P$ is $d_{G}(H_2C) + d_{G}(CP)$.
We choose the smaller proximity from $H_2$ to $P$, and compute the density field using that value.
\section{Domain Problem and Data Preparation}
\label{sec:proc}

% We extract the road network from OpenStreetMap (OSM)\footnote{\href{https://www.openstreetmap.org/}{https://www.openstreetmap.org/}} to construct our map service.  We acquire the traffic data from the open source data platform of Shenzhen\footnote{\href{https://opendata.sz.gov.cn/data/dataSet/toDataDetails/29200\_00403590}{https://opendata.sz.gov.cn/data/dataSet/toDataDetails/29200\_00403590}}. In addition, we use hospitals in Futian district as the hotspots, whose locations can be acquired from the internet. The evaluation index of accessibility of our case studies is the access time from the hospital to the target point.

This section first introduces the domain problem of accessibility measurement for public service facilities (Sec.~\ref{ssec:accessibility}).
Next, we introduce the data used in the study, and describe the data preparation for topology density map construction (Sec.~\ref{ssec:data}). 
% Then, we give the definition of \emph{Accessibility} which is important to our work.

%%%%%%%%%%%%%%%%%%%%%%%%%%%%%%%%%%%%%%%
%%%%%%%%%%%%%%%%%%%%%%%%%%%%%%%%%%%%%%%
\subsection{Accessibility}
\label{ssec:accessibility}

Providing easy access to public service facilities has always been a high-priority topic in urban planning.
Here, each public service facility is regarded as a POI.
POI accessibility is represented as scalar fields distributed over space, which can be visualized using a density map.
% The impact from a POI is a scalar field with abstract concept.
% The distribution of the impact from the POIs can not be regarded the same as the population density distribution which is the number of target objects in one unit.
Short et al.~\cite{short2010nonlinear} pointed out that accessibility in an urban area is better computed as access time rather than physical distance.
Furthermore, commuting by vehicles is the most common travel mode in cities.
Hence, many studies (\emph{e.g.},~\cite{scheepens_2012_interactive, jin2019evaluating}) used vehicle access time to indicate POI accessibility.
% In previous studies, Scheepens et al. used access time to define the potential spatial accessibility from one location to the hospital in the rural area. Jin et al.~\cite{} used distance to measure the spatial accessibility which can find the scope of the service of public medical service facilities and can find the distribution balance of public medical service facility. Both of these methods are highly related to the access time for vehicle on the road. 

%  Furthermore, in POI analysis because most of the actions need to take the vehicles, so in our study, we can think that to some extent the impact of a POI is closely related to the access time for vehicles on the road. 
This work also employs vehicle access time on roads as an index to indicate POI accessibility.
A long access time between a source POI and a target location indicates low POI accessibility, and vice versa. 
% Then we define accessibility as the ability of the POI accessing to the target point.
% In reality, we can only compute an average traveling speed for each road segment, and then measure access times from a POI to every intersection in a road network.
As illustrated in Fig.~\ref{fig:uniqueness}, the path from POI $H$ to location $P$ can be modeled as $n$ road segments $\{seg_i\}_{i=1}^n$ and walkway $w$ between $P$ and a surrounding intersection.
% We use the reciprocal of the access time to calculate the accessibility from a POI to each intersection.
Hence we can compute POI accessibility $Acc$ of $H$ at location $P$ as:
\begin{equation}
Acc(P, H)=
    (\sum\limits_{i = 1}^{n} t(seg_i) + t(w)) ^{-\alpha} %\eqno{(3)}
\label{equ:acc}    
\end{equation}
where $t(\cdot)$ denotes the vehicle access time on a road segment $seg_i$ or the travel time spent on walkway $w$, and $\alpha$ is the accessibility attenuation coefficient that measures the accessibility decay speed.
Peeters and Thomas~\cite{peeters_2000_distance} summarized that $\alpha$ typically lies in the range [0.9, 2.29], where small values are preferred for large accessibility ranges, while large values are chosen for small accessibility ranges.
This work adopts a small attenuation coefficient of 1 for large accessibility ranges.
Notice also that $t(seg_i)$ is a time-variant function that is dependent on the dynamic traffic conditions of road segment $seg_i$.

\begin{figure}[t]
  \centering
  \includegraphics[width=\linewidth]{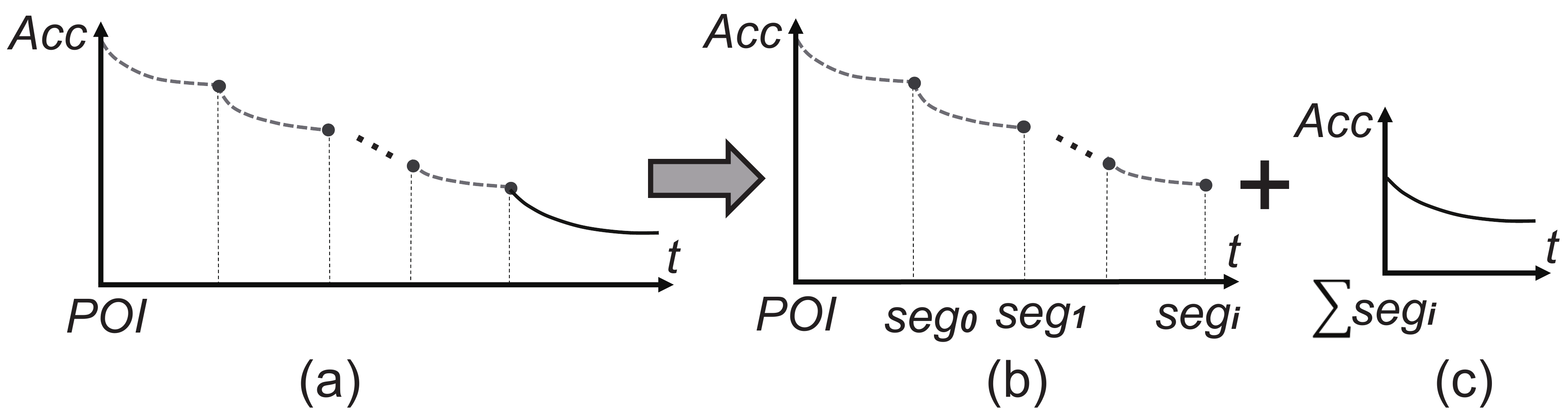}
  \vspace{-2em}
  \caption{The overall accessibility (a) between a POI and a location can be divided into (b) accessibility measured on road segments and (c) accessibility measured on walkway from intersection to the location.
  }~\label{fig:density distribution}
  \vspace{-2.5em}
\end{figure}

Fig.~\ref{fig:density distribution} illustrates the relationship between accessibility ($y$-axis) and access time ($x$-axis).
The overall accessibility (Fig.~\ref{fig:density distribution}(a)) can be decomposed into two parts:
the accessibility measured on road segments (Fig.~\ref{fig:density distribution}(b)), and the accessibility measured on a walkway from a surrounding intersection to an arbitrary location in the region (Fig.~\ref{fig:density distribution}(c)).
% The $y$ axis is the accessibility (impact) and the $x$ axis is the search distance which is the network distance between two intersections on the road.  
% There is a threshold for the search distance. 
% If the search distance exceeds the threshold, we think that this intersection is no longer affected by the current emergency response agency. Furthermore, this threshold depends on the access time from the emergency response agency to the target intersection. If the access time exceeds a certain value, then we think that the current emergency response agency has no or infinite impact on this intersection. 
% There is a threshold for the search distance determined by the access time from the emergency response agency to the target intersection. If the access time exceeds a certain value, then we think that the current emergency response agency has no impact on this intersection. Thus, if the 
Given the nature of the traffic data used in the work, we can only compute the average vehicle speed for each road segment.
In consideration of this fact, we only compute the accessibility value for each intersection, but not for any point on a road segment.
In contrast, we can compute the accessibility value for any point along the walkway.
Hence, we use dash lines for road segment accessibility and solid line for walkway accessibility.
Notice that both road segment and walkway accessibilities are inversely proportional functions of access time.
The function slopes are determined by accessibility attenuation coefficient $\alpha$, which we set to 1 for a gradual decay effect.
% segment represents the edge of the graph and each intersection represents the vertex of the graph (The details are shown in 4.2). 

% Therefore, we can get the set of all segments $Seg = \left\{\ Seg_0, Seg_1,……,Seg_n\right\}$.

\begin{figure*}[h]
\centering
  \includegraphics[width=\linewidth]{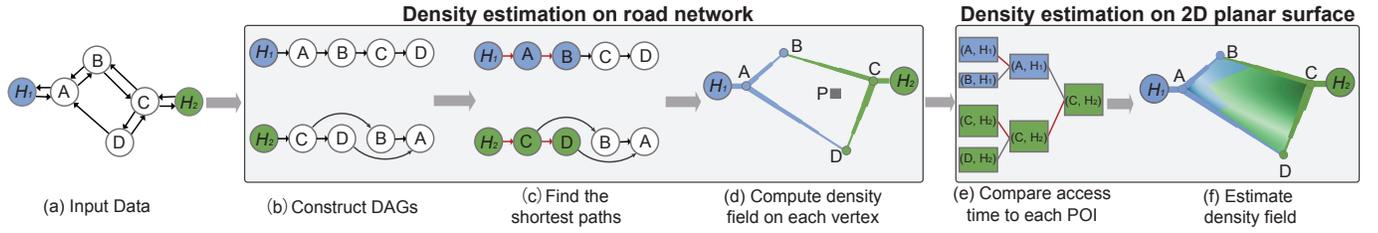}
  \vspace{-1em}
  \caption{Topology density map construction: (a) An example of inputs with $H_1$ and $H_2$ as POIs, and a road network with vertices $A$, $B$, $C$, $D$. (b) For each POI, we first construct a DAG that starts from the POI and connects all network vertices. (c) We can easily find the shortest path from the POI to all vertices with the DAG. (d) We can then compute the density field for each vertex. (d) Next, we extend the density estimation from road network to 2D planar surface. (e) Finally, we compute density fields for all points in the region and construct a topology density map.
}
  \label{fig:medthodology}
  % \vspace{-0.1cm}
\end{figure*}

\subsection{Data Preparation}
\label{ssec:data}
We illustrate the applicability of topology density map by visualizing and analyzing the accessibility for hospitals in Shenzhen, China.

% \noindent
\textbf{Data Collection.}
Our topology density map is constructed from three datasets:
i) We extract geographical data and the road network from the open street map (OSM)\footnote{\href{https://www.openstreetmap.org/}{https://www.openstreetmap.org/}}, which is also utilized for our map visualization.
ii) We acquire traffic data from an open data platform\footnote{\href{https://opendata.sz.gov.cn/data/dataSet/toDataDetails/29200\_00403590}{https://opendata.sz.gov.cn/data/dataSet/toDataDetails/29200\_00403590}} of Shenzhen, China.
We deduce dynamic traffic conditions from the traffic data.
iii) We retrieve hospital locations from the Baidu Map Service\footnote{\href{https://api.map.baidu.com/lbsapi/getpoint/index.html}{https://api.map.baidu.com/lbsapi/getpoint/index.html}}.
The hospitals are POIs analyzed in case studies.
% We also use Baidu Map Open API\footnote{\href{https://lbsyun.baidu.com/}{https://lbsyun.baidu.com/}} to collect the coordinates~(longitude and latitude) of road intersections. \\

% \noindent
\textbf{Road Segmentation.}
We define road segment as a part of the road between two adjacent intersections. 
% data we get from Baidu Map Open API includes a sequence of which are the start and end of each segment.  
We treat each road segment, instead of the road (represented as ``way" in OSM), as the unit of accessibility analysis in this study.
% Roads are usually not parallel and intersect to produce intersections between each other.
This introduces several advantages:
First, by dividing a long road into several road segments, we can deduce more accurate access times because different road segments typically share different speeds.
% as each part of a road between two intersections has different speed, defining a segment containing several intersections leads to a less accurate access time estimation.
% The accuracy of the computation will be affected, if we consider a long road which has several segments as a single unit. 
Second, we can easily establish a DAG, in which an edge represents a road segment and a vertex represents an intersection.
To accomplish road segmentation, we find all intersections in a road network, and group a sequence of edges connecting those vertices as one unit corresponding to a road segment.
By combining road segments with traffic data, we can generate an average speed for each road segment in a certain time period.

% \noindent
\textbf{POI-Intersection Connection.}
To study how POI accessibility propagates along a road network, we need to connect POIs to road network.
% In our study, we see POIs as the source of the impact, the impact will propagate along the road network. 
POIs have three possible locations relative to a road network in an urban area: right on an intersection, just beside a road, or neither beside any road nor on any intersection. 
In the first situation, we can simply treat a POI as a vertex in the road network.
For the other situations, Weng et al.~\cite{weng2018homefinder} created a new vertex that is the projection point of a POI to its nearest road in the road network.
% left an POI at its location if it is just beside a road, or projected it to the road that is closest to its location.
In contrast, we connect a POI to its nearest intersection.
We do this for two reasons.
First, due to the availability of input data, traffic conditions and road topology in the planar surface enclosed by a road network are unknown to us.
A marginal difference exists between projecting a POI to its nearest road and connecting it directly to the nearest intersection. 
Second, we can only measure the average speed of each road segment, whilst a vehicle speed is non-uniform when driving on a road.
Hence, we cannot measure the exact access time from the projection point to the corresponding intersection.

% \noindent

% \zw{keep consistent with nodes or vertices.}
% so the direction of each edge is the same as that of the corresponding road, if the number of edge between two adjacent vertices is two and with different directions, it represents a two-way road, if the number of it is one, it represents only one direction to pass, if there is no edge connection, it represents that the two vertices are not accessible. \textcolor{red}{whether add a group of figures} Each vertex is stored by the accessibility (will be mentioned in section 5.1) and the weight of the edge is the access time from one vertex to another.
\section{Topology Density Map}
\label{sec:method}

In this section, we introduce the key steps on how to construct a topology density map:
first, to estimate the density fields on road network (Sec.~\ref{ssec:estimate_road}), then extend density field estimation to a 2D planar surface (Sec.~\ref{ssec:estimate_surface}).
The construction process is illustrated in Fig.~\ref{fig:medthodology}, where the inputs (Fig.~\ref{fig:medthodology}(a)) include POIs $H_1$ \& $H_2$, and road network $G$ with vertices $\{A, B, C, D\}$.
The goal is to estimate a density field for arbitrary point $P$ within polygon ABCD.
Next, we present the visual effects of the topology density map (Sec.~\ref{ssec:visual_effects}), and its advantage over NKDE in terms of intuitiveness (Sec.~\ref{ssec:comp_NKDE}).

% In this study, we propose a new method to draw the density map in the urban area by analyzing the density distribution of the emergency response agency impact.
% \subsection{Topology Density Estimation}
% Before \textcolor{blue}{constructing} the topology density map, we should estimate the density distributed on the research region. In our method, we estimate the density in the form of accessibility. The estimation of the density includes two steps: estimating the density on the road based on network distance and estimating the density on the planar surface based on the Euclidean distance~(Fig.~\ref{fig:medthodology}).  Fig.~\ref{fig:density distribution}(a) shows the overall trend of density distribution for topology density map. Fig.~\ref{fig:density distribution}(b) presents the density distributed on the road and Fig.~\ref{fig:density distribution}(c) presents the density distributed on the planar surface.
% The two steps have different method to compute: the density one is based on network distance and another is based on the Euclidean distance.
             
\subsection{Density Estimation on Road Network}
\label{ssec:estimate_road}
In this step, we estimate the density propagation along the road network from POIs, as follows:

\vspace{1mm}
\textbf{Construct DAGs (Fig.~\ref{fig:medthodology}(b)).}
After retrieving the access times for the road segments and connecting the POIs to the road network, we construct a DAG for each POI to facilitate the accessibility propagation estimation.
Each DAG starts from the corresponding POI, and connects all intersections by directional road segments.
Here, vertices correspond to intersections, and edges correspond to road segments. %, with edge weights indicating access times.
We also consider the road direction, so the edge direction indicates the travel direction of the corresponding road segment.
Fig.~\ref{fig:medthodology}(b) presents two example DAGs for POIs $H_1$ and $H_2$ in blue and green, respectively.
Notice that $H_1$ is first connected to vertex $A$ as the nearest intersection, while $H_2$ is first connected to vertex $C$ as the nearest intersection.
All vertices in the road network are eventually included in the DAGs, with differences in vertex orders that reflect the road topology.

%  connects all nodes 
% The input of topology density map is a DAG mentioned in Section \ref{data}.
% The vertices of the graph represent the intersections in the road network and the edges of the graph are the directional roads.
% Fig.~\ref{fig:medthodology}(a) is an example DAG of the road network, vertices $H_1$ and $H_2$ with color are two different POIs.
% The other vertices are the intersections in the road network and the point $p$ is a target point which we want to estimate the density there.
% In this example, we default that the road \textcolor{blue}{in Fig.~\ref{fig:medthodology} is unidirectional.
% In reality, most of the roads in the city are bi-directional.
% However, when considering the density in 2D planar surface, we only consider the roads enclose the 2D planar surface, the roads not close to the 2D planar surface will not generate any influence in our method.
% Therefore, in order to show the methodology more clearly, we only draw unidirectional roads instead of bi-directional roads which are most of the actual situation}.

\vspace{1mm}
\textbf{Find the shortest paths (Fig.~\ref{fig:medthodology}(c)):}
Given a DAG, we can easily find the shortest paths from a POI to all vertices using the Dijkstra algorithm.
All edges here are weighted according to the access time retrieved from the traffic data.
For each vertex, we select the POI with the least access time as the most accessible POI.
This has practical significance, as people typically choose the most accessible POI and ignore others in reality.
% Here we denote the easiest accessible POI for an intersection $v$ as $P_v$.
Fig.~\ref{fig:medthodology}(b) indicates that the most accessible POI for $A$ and $B$ is $H_1$, while that for $C$ and $D$ is $H_2$.
% Hence we can derive $P_{A} = P_{B} = H_1$, while $P_{C} = P_{D} = H_2$. 

\vspace{1mm}
\textbf{Compute density field on each vertex (Fig.~\ref{fig:medthodology}(d)):}
After identifying the most accessible POI for each intersection, we can then compute the density propagation along the road network.
% The density for each intersection is computed in the form of accessibility from its easiest accessible POI as Equation~\ref{equ:acc}.
Here, we use the color of the most accessible POI to encode an intersection, that is, $A$ \& $B$ are colored blue, while $C$ \& $D$ are colored green.
Moreover, to better depict density propagation process on the road network, we further encode road segments with tapered lines that are preferred for graph readability~\cite{holten_2011_entended}.
Here, a line is colored according to its connecting vertex with a high accessibility value, and the line width gradually decreases from the vertex with a higher accessibility value to the other vertex with a lower accessibility value.
Taking road segment $BC$ for example, the line color is green for vertex $C$, and the line width is decreases from $C$ to $B$.
Users can easily deduce that vertex $C$ has a higher accessibility to $H_2$ than vertex $B$'s accessibility to $H_1$.

% Since we only consider the greatest impact on intersections, each intersection is only affected from the same POI in the same direction. \zw{do not understand}
% Therefore, the density of POIs on the road is encoded by lines with color, the size of density is encoded by the thickness of lines. The density of the road between the two adjacent intersections is determined by the two intersections themselves. Due to our previous hypothesis is that each segment is uniform, the density on each segment is uniformly reduced from the intersection with higher accessibility to the intersection with lower accessibility, and the color of the line is the same as the intersection with higher accessibility. 

\subsection{Density Estimation on 2D Planar Surface}
\label{ssec:estimate_surface}

In this step, we extend the density estimation from a 1D road network to a 2D planar surface to fulfill the intuitiveness requirement.
Here, we illustrate the process with density estimation for an arbitrary point $P$ as in Fig.~\ref{fig:medthodology}(d).
Due to the lack of road network and traffic condition in area $ABCD$, we adopt a simplified access time measurement in a linear correlation with the Euclidean distance between $P$ and the intersections.
Here, we adopt a twofold process as follows:

\vspace{1mm}
\textbf{Compare access time to each POI (Fig.~\ref{fig:medthodology}(e)):}
As discussed in Sec.~\ref{ssec:data}, we associate a POI to the road network by connecting the POI to the nearest intersection.
Similarly, we also associate an arbitrary point $P$ to road network by connecting $P$ to the neighboring intersections, \emph{i.e.}, $\{A, B, C, D\}$, yielding straight walkways $AP$, $BP$, $CP$, \& $DP$.
Given that we have found the most accessible POI for each intersection, we can compute the access time from $P$ to each POI by summarizing the network proximity of each road segment and the Euclidean distance of the straight connection.
Taking the access time between $P$ and $H_1$ for an example, we have two options:
1) to connect $P$ to $A$, yielding an access time of $t(H_1A) + t(AP)$;
2) to connect $P$ to $B$, yielding an access time of $t(H_1A) + t(AB) + t(BP)$.
Given that the access time on the road network is typically shorter than that on walkways (travel by vehicle \emph{vs.} by walking), we select option 1.
Similarly, we can also compute the access time between $P$ and $H_2$ as $t(H_2C) + t(CP)$.
The comparison of the access times to $H_1$ and $H_2$ indicates that the access time to $H_2$ is lower, and hence the accessibility to $H_2$ is higher.

\vspace{1mm}
\textbf{Estimate density field (Fig.~\ref{fig:medthodology}(f)):}
Hence, we can estimate a density field for $P$ with respect to $H_2$ as:

\begin{equation}
\lambda(P | H_2, G) = \frac{1}{r}K(\frac{(t(H_2C) + t(CP))^{-\alpha}}{r})
\label{equ:density_final}
\end{equation}
where $\lambda(P | H_2, G)$ denotes the density field at $P$ given POI $H_2$ and road network $G$, $t(H_2C) + t(CP))^{-\alpha}$ reflects the $H_2$ accessibility at location $P$ (Equation~\ref{equ:acc}), and $K$ is the kernel function with $r$ as the kernel radius (Equation~\ref{equ:nkde}).
The color of the point is assigned in accordance with the color of the source POI, \emph{i.e.}, $H_2$.
Similarly, we can estimate the density fields for other points in the 2D planar surface.
Then, we render each pixel corresponding to its density field and the source POI to generate a topology density map.
An example of the generated topology density map is presented in Fig.~\ref{fig:medthodology}(f).

\begin{figure}[t]
  \centering
  \includegraphics[width=\linewidth]{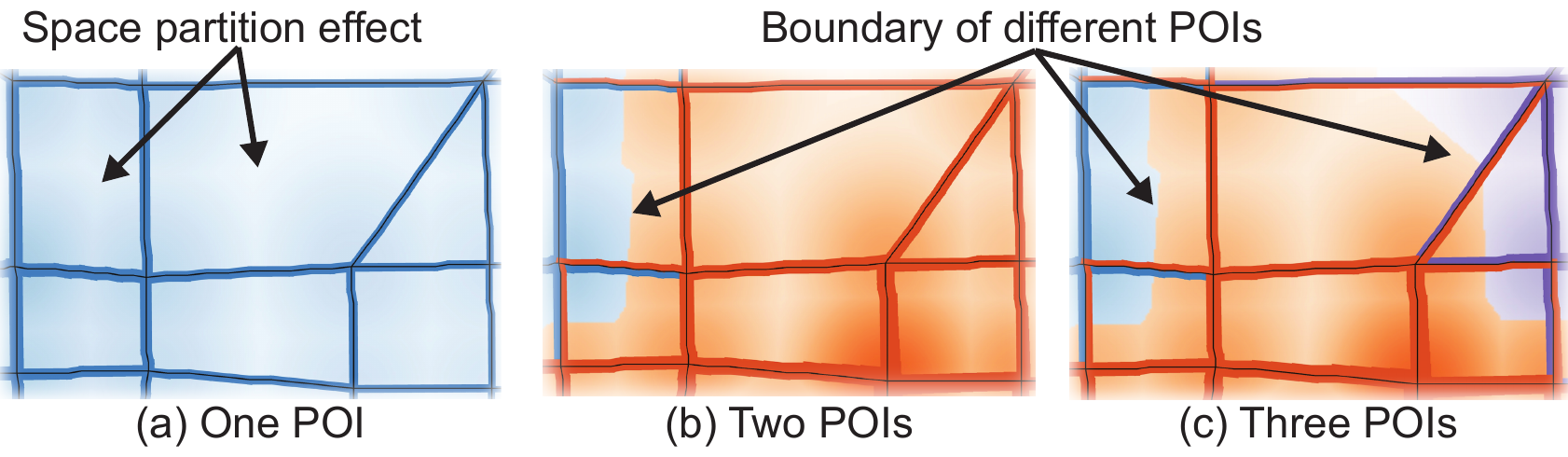}
  \vspace{-1em}
  \caption{Visual effects of topology density map: 
   color values clearly reveal quantitative density fields distributed in the space when there is only one POI (a);
   and additional color hues depict accessibility to different POIs when there are two (b) and three (c) POIs.
  % the selected region is affected by two POIs, the boundary of different POIs is obvious and one planar surface can be affected by different POIs simultaneously;
  % and (c) the selected region is affected by three POIs.
  }~\label{fig:space_partition}
  ~\vspace{-1em}
\end{figure}

%%%%%%%%%%%%%%%%%%%%%%%%%%%%%%%%%%%%%%%%%%%%%%%%%%%%
%%%%%%%%%%%%%%%%%%%%%%%%%%%%%%%%%%%%%%%%%%%%%%%%%%%%
\subsection{Visual Effects of Topology Density Map}
\label{ssec:visual_effects}

Topology density map adopts color hues to encode different POIs, and color values to indicate quantitative density fields.
In this way, users can not only distinguish which POI is the easiest accessible, but also how much easy to access the POI from arbitrary locations in a 2D planar space.
% In addition, it can also reveal the reasons for multiple POIs affecting the 2D planar surface. \zw{dont understand}
% The method generates a weighted Voronoi diagram like space partition effect where the boundary between clusters are the boundary of different range from each intersection.
In each region segmented by road segments, the method generates a visual effect corresponding to space partition by a weighted Voronoi diagram where the centers are intersections and the weights are network-based proximity to POIs; see Fig.~\ref{fig:space_partition}(a) for an example.
Users can easily identify which intersection is the easiest accessible from a location in the region.
% the range of each cluster indicates that it is most affected by the corresponding intersection in this cluster. 

Fig.~\ref{fig:space_partition}(a-c) shows visual effects of topology density map with of one to three POIs, respectively.
There is only one POI in Fig.~\ref{fig:space_partition}(a), so all density fields are colored in blue.
Meanwhile, one can notice that colors are gradually fading from intersections to region centers, indicating those locations near by intersections are easier accessible.
When more than one POIs are presented, there will be multiple color hues in the topology density map, \emph{e.g.}, two hues for two POIs in Fig.~\ref{fig:space_partition}(b), and three hues for three POIs in Fig.~\ref{fig:space_partition}(c).
One can easily identify boundaries between different color hues, separating the space into regions by the easiest accessible POIs.
Here we can notice that the region is mostly affected by the orange POI.
% The different colors in the region represent that the whole region is affected by different POIs.
Notice that the space partition effect and boundaries between two POI regions may be affected by various parameters when constructing topology density map, \emph{e.g.}, bandwidth and kernel function.
The parameter settings will be further discussed in Sec.~\ref{ssec:parameter_selection}. 
% The density map generated by our method will not give users specific and accurate values or any clear suggestions.
% Although the cluster size will be affected by different combinations of kernel and bandwidth, we use the same combination to generate the same density map.
% The space partition shown by our method can provide users with such information as which POI this region is affected by, the proportion of different POIs affecting in the same region and the proportion of each intersection affecting the corresponding region.
% As for the specific judgment of the POI analysis results, it is up to the user to make the final decision based on their own empirical experience.
% One of functions for topology density map is to assist users on their decision making.

\begin{figure}[h]
  \centering
  \includegraphics[width=\linewidth]{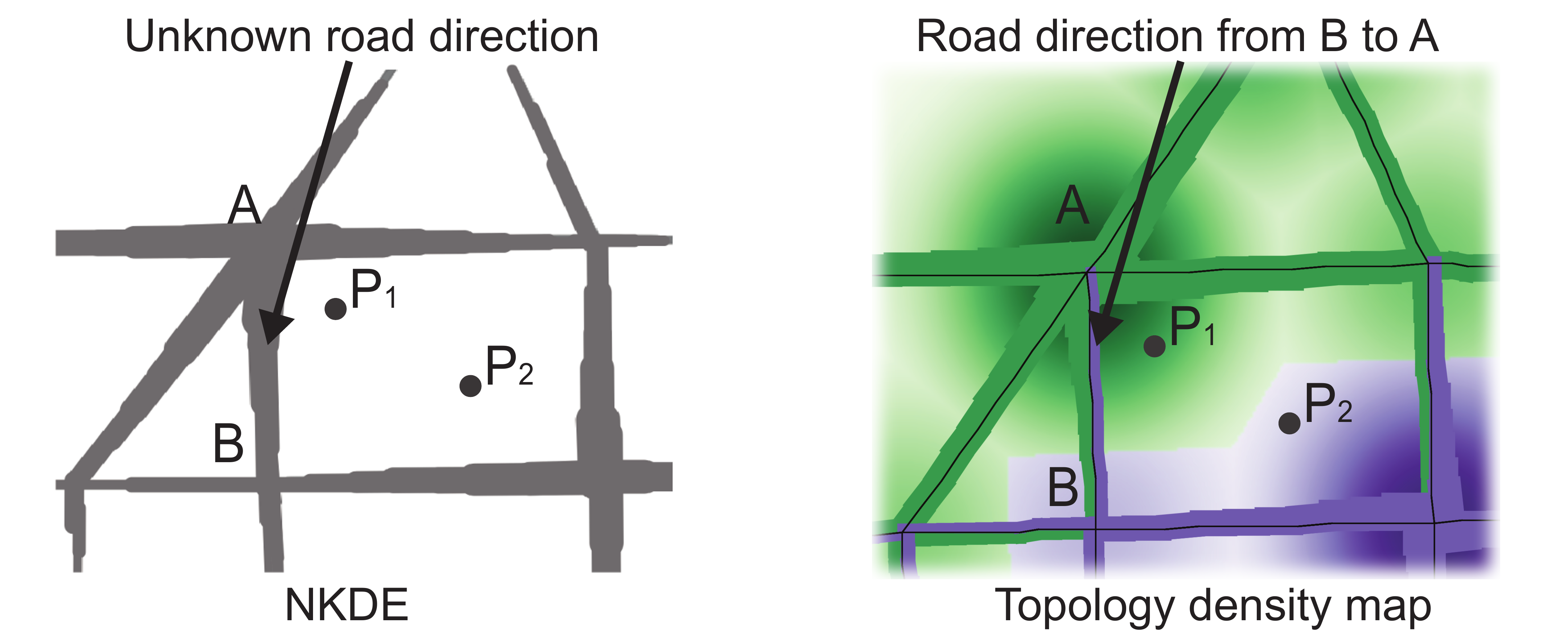}
  \vspace{-1em}
  \caption{   
  Comparing topology density map (right) with NKDE (left).
  Topology density map is more intuitive than NKDE in presenting density distribution over 2D space and depicting road directions.
  }
  \vspace{-1em}
  ~\label{fig:NKDE_comp}
\end{figure}

%%%%%%%%%%%%%%%%%%%%%%%%%%%%%%%%%%%%%%%%%%%%%%%%%%%%
%%%%%%%%%%%%%%%%%%%%%%%%%%%%%%%%%%%%%%%%%%%%%%%%%%%%
\subsection{Comparison with NKDE}
\label{ssec:comp_NKDE}

Planar KDE and NKDE are state-of-the-art approaches for density estimation.
However, planar KDE is inaccurate for urban data analysis and visualization because it omits the road network.
In contrast, NKDE estimates density based on the linear propagation along a 1D road network, which has been applied to analyze taxi pick-ups/drop-offs~\cite{xia2019identify, pei2015density} and traffic accidents~\cite{harirforoush2019new}.
Hence we only compare our topology density map with NKDE in terms of intuitiveness (R2). 

% NKDE also introduces the road topology into the method of density map. However, NKDE focuses on linear data and it just estimates the density on the 1D road which cannot cover the whole region in urban area.  They analyze urban POIs by counting the number of traffic trajectories or accidents at one location on the road over a period of time. We pay more attention on the traffic condition and generate density map by using nonlinear data.

Fig.~\ref{fig:NKDE_comp} presents a comparison of the visualizations generated by NKDE (Fig.~\ref{fig:NKDE_comp}(left)) and our topology density map (Fig.~\ref{fig:NKDE_comp}(right)).
Both visualizations are built from the same POIs  and local structure of a road network.
Here, our topology density map reveals the density fields over the entire space, while NKDE only depicts the density fields along the roads.
For instance, the density values for $P_1$ \& $P_2$ are unknown in Fig.~\ref{fig:NKDE_comp}(left).
In contrast, the most accessible POI for $P_1$ is the green POI, while that for $P_2$ is the purple POI; and density value for $P_1$ is higher than that for $P_2$, from Fig.~\ref{fig:NKDE_comp}(right).
In this sense, our topology density map supports better the functions of \emph{lookup} and \emph{comparison} than NKDE.
Moreover, our topology density map also depicts directions of road segments using tapered lines, whilst NKDE only presents density fields along road segments without direction information.
Taking the road segments between intersections $A$ and $B$ in Fig.~\ref{fig:NKDE_comp}(right) for an example, $A$ is more accessible to the green POI since $A$ is colored green, while $B$ is more accessible to the purple POI.
One can also figure out that the road segment on the right side is from $B$ to $A$, while the one on the left side is from $A$ to $B$, by checking either colors of the road segments or taper directions.

In summary, our topology density map surpasses the performance of NKDE in presenting the density fields over 2D space and depicting road directions.

% In Fig.~\ref{fig:NKDE and topology density map}(b), when we consider the direction of the road there may exist such situation like in the figure, the intersection $A$ and $B$ are affected by two different POIs, but the color of this road on this direction is the same as that of its starting point.
% This is because, in this study, we consider the average traffic conditions between intersections, but the specific speed of each point on the road is unknown, so we just consider the source of the selected road segment the same as its start intersection.
\section{Case Studies and Expert Interview}
\label{sec:case_study}

\begin{figure*}[t]
  \centering
  \includegraphics[width=0.995\linewidth]{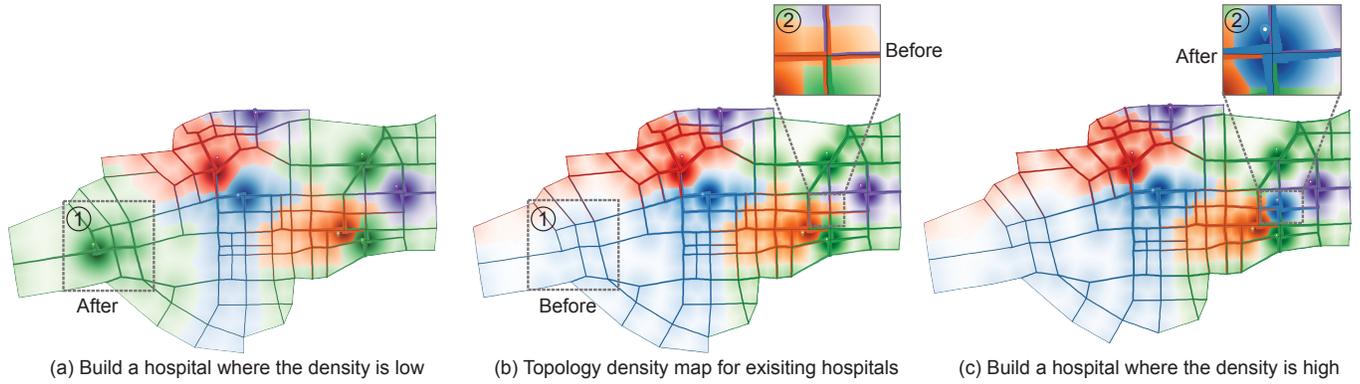}
  \vspace{-1em}
  \caption{
  Topology density map built from existing hospitals (b), after adding a new hospital in region 1 with low density (a), and after adding a hospital in region 2 with high density (c).
 Region 1 is a preferred location, as density field is more evenly distributed in (a) than in (c).
  }~\label{fig:study2}
  \vspace{-1em}
\end{figure*}

We conduct two case studies to verify whether the topology density map satisfies the requirements discussed in Sec.~\ref{ssec:requirement}.
We select the Futian district in Shenzhen, China as the study area.
The area is presented as the main view in Fig.~\ref{fig:teaser}.
Feedbacks from three independent domain experts are presented at the end. 
% We extract the road network from OpenStreetMap (OSM)\footnote{\href{https://www.openstreetmap.org/}{https://www.openstreetmap.org/}} to construct our map service.  We acquire the traffic data from the open source data platform of Shenzhen\footnote{\href{https://opendata.sz.gov.cn/data/dataSet/toDataDetails/29200\_00403590}{https://opendata.sz.gov.cn/data/dataSet/toDataDetails/29200\_00403590}}. In addition, we use hospitals in Futian district as the POIs, whose locations can be acquired from the internet. The evaluation index of accessibility of our case studies is the access time from the hospital to the target point.

\subsection{Study 1: Analyzing the Influence of Road Topology and Traffic Conditions on Accessibility}
\label{ssec:study1}

We first verify whether the topology density map can present the density fields accurately (requirement R1 as in Sec.~\ref{ssec:requirement}), by taking both road topology and traffic conditions into consideration.

\vspace{1mm}
\noindent
\textbf{Road Topology}.
\fzz{We first zoom in to the selected region shown in Fig.~\ref{fig:teaser}(a).
The two different colors indicate that this region is easily accessible to two POIs denoted as $H_1$~(red color) and $H_2$~(blue color).
In Fig.~\ref{fig:teaser}(a), Regions $1$~($BCDE$) and $2$~($ABEF$) are easily accessible to both $H_1$ and $H_2$, while Region $3$~($DFGI$) is fully covered by POI $H_2$. 
From the density distribution along the road segments, we can find reasons for the phenomena:
intersections $A$, $B$, and $C$ are easily accessible to POI $H_1$, while intersections $D$, $E$, and $F$ are easily accessible to POI $H_2$.
Moreover, $H_2$ propagates along paths $F$-$E$-$B$ and $E$-$D$-$C$.
To explore the influence of road topology, we generate a new density map by removing paths $F$-$E$ and $E$-$D$.
Fig.~\ref{fig:teaser}(b) shows the result, which is the same region as that in Fig.~\ref{fig:teaser}(a).
The comparison of Figs.~\ref{fig:teaser}(b) with~\ref{fig:teaser}(a) indicates the changes in density field distribution. 
Region $1$, which is easily accessible to both $H_1$ and $H_2$ in Fig.~\ref{fig:teaser}(a), is now fully covered by POI $H_1$ in Fig.~\ref{fig:teaser}(b).
In contrast, Region $3$, which is previously only covered by POI $H_2$ in Fig.~\ref{fig:teaser}(a), is now accessible to both $H_1$ and $H_2$ in Fig.~\ref{fig:teaser}(b).}
% Section 5.1 mentioned that the impact from a POI is first transmitted to each intersection through the shortest path on network distance in the form of accessibility, and then propagated to the whole corresponding planar regions from the intersections.

Removing paths $F$-$E$ and $E$-$D$ makes path $F$-$E$-$D$ inaccessible, yielding changes to road topology.
Therefore, the previous path $H_2$-$F$-$E$-$D$ with the shortest access time is no longer available, changing the most accessible POI at intersection $D$ from $H_2$ to $H_1$, and that at intersection $E$ from $H_2$ to $H_1$.
This affects the construction of the topology density map in terms of estimating the density propagation from POIs and determining the source POI (Sec.~\ref{ssec:estimate_surface}).
Hence, the resulting topology density map can also reflect the changes.
% For instance, region $1$ in Fig.~\ref{fig:teaser}(b) is fully covered by POI $H_2$, while region $3$ in fig.~\ref{fig:teaser}(b) is affected by both POI $H_1$ and POI $H_2$.
% compared with Fig.1(a), in Fig.1(b) the POI affecting intersection $C$ changes from $H_2$ whose shortest path is $H_2$-$E$-$D$-$C$ to $H_1$ whose shortest path is $H_1$-$A’$-$B’$-$C’$. 
In contrast, Euclidean distances between arbitrary locations in the regions and POIs are unaffected by road topology change, resulting in unchanged planar KDE.
% This case shows that although the Euclidean distance between points on the map does not change, the change of the road topology leads to the change of network distance on the road, resulting in the change of density distribution and POIs' range.

\vspace{1mm}
\noindent
\textbf{Traffic Condition}.

We then zoom in to the selected region shown in Fig.~\ref{fig:teaser}(c).
The figure illustrates that the region is easily accessible to two POIs denoted as $H_3$~(orange color) and $H_4$~(green color).
Region $4$ is accessible to both $H_3$ and $H_4$, while Region $5$ is fully covered by POI $H_4$.
In general, the density field distribution shows a trend of continuous accessibility propagation toward the outbound of the district.
However, when focusing on the green region ($H_4$), we notice some unusual patterns.
A very narrow green part can be observed in Region $4$, while the green part increases again in Region $5$.
% , rather than this selected region in Fig.\ref{fig:teaser}$(c)$ which one POI generate more than one clusters and each cluster connected by narrow sections.
The variation of the density fields along the road segments reveals that path $M$-$N$-$O$ connects Regions $4$ and $5$. We obtain the average speeds of these road segments ($M$-$N$ and $N$-$O$), and find that these segments exhibit high average speeds, reducing the access time through path $M$-$N$-$O$ even though the Euclidean distance is long.

To explore the influence of traffic conditions on density fields, we reduce the average speeds of road segments $M$-$N$ and $N$-$O$ manually.
The density field changes correspondingly, as illustrated in Fig.~\ref{fig:teaser}(d).
The easiest accessible POI in Region $4$ shows no change.
However, it changes from $H_4$ to $H_3$ in Region $5$ and the density fields decrease as well, caused by lower speeds of road segments $M$-$N$ and $N$-$O$. 
% Since the road structure and average speed of roads around region $4’$ have not changed, the density of region $4’$ and the range of the POI will not change in this region. 
% Due to the lower average speed of the road segments $K$-$L$ and $L$-$M$ from east to west, the POI with the shortest access time to target locations in region $5’$ changes from POI $H_4$ to POI $H_3$. 
% Because of the change of the shortest path of the network distance, the density of each intersection in region $5$ also changes. And some intersections previously affected by $H_4$ are now affected by $H_3$, so compared with the Fig.\ref{fig:teaser}$(c)$ the density in this region decreases.
This phenomenon indicates that although the Euclidean distances between intersections remain unchanged, dynamic traffic conditions can lead to variations in shortest paths on a road network and ultimately alter the density distribution on a 2D planar surface.

\subsection{Study 2: Finding the Optimal Location of a New Public Service Facility}
\label{ssec:study2}

Next, we demonstrate the intuitiveness of topology density map in presenting density distribution (requirement R2 as in Sec.~\ref{ssec:requirement}), which facilitates the process of finding the optimal location of a new public service facility.
% We then want to verify that topology density map visualize the density distribution intuitively, which covers the L2 mentioned in Section~\ref{limitation}.2.
Here, we use hospitals in Futian District crawled from Baidu Map Service as targeting public service facilities, and access time to measure accessibility to those hospitals.
% We still use hospitals as POIs and use access time to measure the impact of hospitals on surrounding regions.
% In Fig.~\ref{fig:case study 2}(a) we can see seven hospitals on the map whose impact is represented by density.
Fig.~\ref{fig:study2}(b) presents density field of accessibility measured from the hospitals, which is unbalanced 
in the district as currently the hospitals are concentrated in the east part of Futian District.
The government decides to build a new hospital and tries to find a suitable location for the hospital.

% As mentioned in Section~\ref{data}, in order to facilitate calculation and provide a more intuitive visualization, they are all moved to the nearby intersections. 
% As Fig.~\ref{fig:case study 2}(a) shows, the density distribution is unbalanced since most of hospitals concentrate at the east of the whole research area. 
% This phenomena indicates that the medical resources is insufficient in the southwestern Futian district.
% most of the POIs locate in the east of the region thus the density distributed in the eastern regions and northern regions are the densest. 
% However, at the southwestern region, there are no POIs there, in addition no matter any POIs the density distributed at this region is very low. It is obvious from the figure that 
A critical consideration here is that the hospital should make the distribution of medical service more balanced. 
In other words, to maximize the increase of hospital accessibility for the whole district.
Two possible options for the hospital location are available.
\begin{itemize}
\item
First, we can put the new hospital in Region $1$ where density is low.
The topology density map is updated correspondingly and presented in Fig.~\ref{fig:study2}(a). 
Obviously, the density field in the southwest part increases, especially that in Region $1$.
This intuitively depicts how much the region can benefit from a new public service facility.
Moreover, because the density first propagates along the road network, and then propagates to a 2D planar surface, the change in density distribution on the road segments is obvious.
In short, putting the new hospital in Region 1 balances the accessibility to medical service.

\item
Second, we can place the hospital in Region $2$ where the density is already high.
The updated topology density map is shown in Fig.~\ref{fig:study2}(c).
Overall, both densities in the selected region and along the surrounding road segments increase.
However, this arrangement does not improve the southwest part with low density.
Hence, building a new hospital in Region $2$ worsens the unbalanced distribution of medical service.
\end{itemize}

The study clearly demonstrates that the topology density map generates intuitive visualizations that can facilitate problem identification and decision making.
% Users can visualize the impact of POIs in the research area by using Topology Density Map both on the road and on the planar surface.

%%%%%%%%%%%%%%%%%%%%%%%%%%%%%%%%%%%%%%%%%%%%%%
%%%%%%%%%%%%%%%%%%%%%%%%%%%%%%%%%%%%%%%%%%%%%%
\subsection{Expert Interview}
\label{ssec:interview}
To evaluate our topology density map, we interviewed three independent domain experts.
 % to obtain their feedback on our topology density map. 
The first expert (Expert A) is an assistant professor in University S specializing in smart city related research.
The second expert (Expert B) is a Ph.D. candidate in University H whose research focuses on data-driven urban planning.
The third expert (Expert C) is a practitioner working in an urban planning institute.
Before the interview, we first made a brief introduction of our method, and then explained the details of the two case studies.
We collected feedbacks in terms of the feasibility, usability, and effectiveness of our topology density map from the experts in the end.
All experts provided valuable feedbacks based on their own background.
The detailed feedbacks are summarized below. 

 \textbf{Feasibility \& Usability.}
 All three experts agreed that using a density map to reveal accessibility of POIs in an urban area is a good choice.
 Our topology density map can really help in their research and work.
 Expert A mentioned that ``% In my research, I use density map to analyze the impact of POIs in urban area. 
 topology density map takes road topology and traffic conditions into consideration, which fits the reality very well.
 If the method is applied in my research, it will help me a lot on decision making."
 Expert B said that “topology density map is of great research value because it considers road structure, and covers the whole city."
 They all agreed that our method visualizes density field intuitively.
 Expert B commented that ``from the cases you introduced, I can see that topology density map does provide an intuitive visualization, especially in case study 2.
 Through topology density map, I can quickly find that the current distribution of hospitals, that is, accessibility to medical resources, is indeed unbalanced.
 After you manually add a new hospital at two different places, the new density maps show the difference intuitively and definitely help me make a decision.''
 
 \textbf{Effectiveness.}
All experts thought that the topology density map is an effective tool for urban analysis.
Expert C shared her own experience in identifying suitable locations for new public service facilities.
In daily work, her team often need to place a certain number of public service facilities evenly around a few residential blocks, but they do not have an intuitive and easy-to-use tool that could assist them in finding optimal locations.
Currently, the team rely on manual approaches:
they first mark streets that are close to the residential area, then measure distance from the residential area to each street, and finally draw contour lines based on the measured distances. 
The process is rather ineffective, and often leads to unsatisfactory results since the distance measurement omits traffic conditions.
 % \zw{One of the methods she uses now is to acquire the street names by calculating the distance to the residential
 % area and then marking these names on the map.
 % The other is to use contour methods to draw the range.}
 % However, these two methods are not user-friendly.
 Moreover, they also find it is difficult to share the planning with others, and would be impossible for others to make modifications.
 She believes that the topology density map is the right tool they are looking for, not only for finding locations of public service facilities, but also for comparing various planning strategies.  
 % visualize the information intuitively which can help them make decision quickly.

\textbf{Suggestions.} 
The experts also gave fruitful suggestions on how to improve the topology density map.
Both Experts A and C commented that developing our method into a mature software that supports interactive planning would be great.
As potential end users, urban planners typically have little knowledge of programming.
Expert B pointed out that the unavailability of non-uniform travel speeds in a road segment affects the accuracy of our method in density estimation.
She suggested that if we get more detailed data, we can consider connecting POIs to road segments directly, rather than to the nearest intersection because the vehicle speeds on a single road segment can vary dramatically.
% is not transferred into the 2D planar surface through the intersection, but through the entrance of the region, such as the gate of the campus. In addition, 
% Expert A also mentioned the limitations of current data processing.
Expert A also suggested that after solving the problem of insufficient and inaccurate data, we can consider some demographic factors, such as population distribution, which are important elements to be considered in urban analysis.
For instance, in case study 2, a more practical strategy is to consider population distribution over space as well.
\begin{figure}[t]
  \centering
  \includegraphics[width=\linewidth]{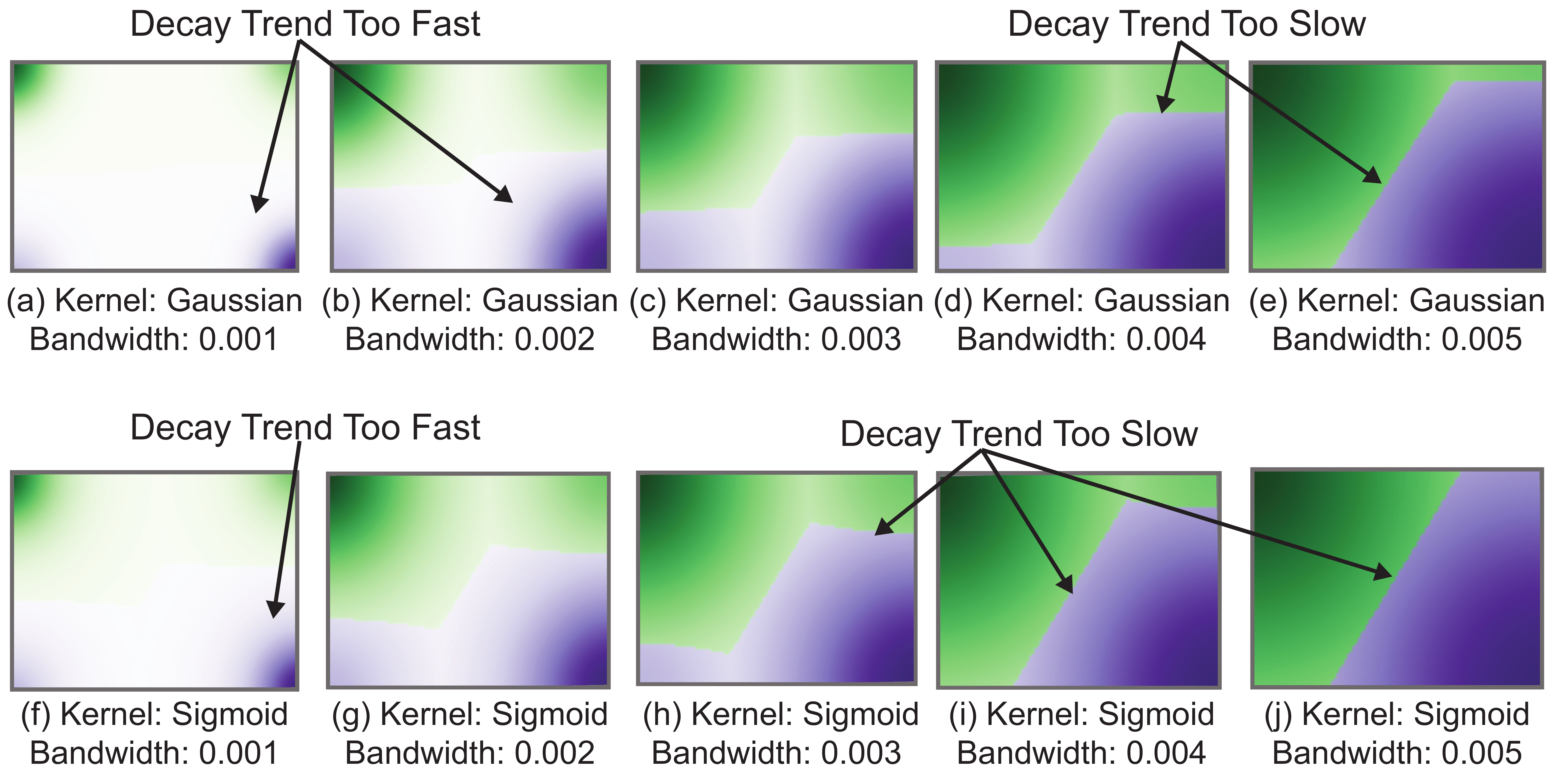}
  \vspace{-2em}
  \caption{Parameter selection: (a) to (e) are Gaussian kernel function with different bandwidths, and (f) to (j) are Sigmoid kernel function with different bandwidths. 
  }~\label{fig:parameter selection}
  \vspace{-2em}
\end{figure}

\section{Discussion}
\label{sec:discuss}

% \begin{table*}[]
% \caption{Running time comparison for topology density map and planar KDE by different number of POIs and image size.}
% \centering
% \begin{tabular}{|l|l|l|l|l|l|l|}
% \hline
% % \multirow{2}*{\diagbox{Size of Image}{Applied Method}{Number of POIs}} 
%         \textbf{Number of POIs}& \multicolumn{2}{c|}{\textbf{5 POIs}}  & \multicolumn{2}{c|}{\textbf{10 POIs}} & \multicolumn{2}{c|}{\textbf{20 POIs}}\\ \cline{1-1}
%  \diagbox {Image Size}{Method}&\multicolumn{1}{c|}{{ \textbf{\begin{tabular}[c]{@{}c@{}}Topology \\density  map\end{tabular}}}}& { \textbf{Planar KDE}} &\multicolumn{1}{c|}{{ \textbf{\begin{tabular}[c]{@{}c@{}}Topology \\density  map\end{tabular}}}}& { \textbf{Planar KDE}} & \multicolumn{1}{c|}{{ \textbf{\begin{tabular}[c]{@{}c@{}}Topology\\ density map\end{tabular}}}}& { \textbf{Planar KDE}} \\ \hline\hline
% 1920*1200& {1403.12ms} & {1028.20ms}& {1351.76ms} & {1083.43ms}& { 1411.33ms}& {1161.22ms}\\ \hline
% 1280*800 & {933.26ms} & {845.42ms}& {954.28ms} & {787.21ms}& {950.34ms} & {802.68ms} \\ \hline
% 640*400& {396.02ms} & {361.67ms}& {432.33ms}& {384.54ms} & {478.29ms}& {364.62ms}\\ \hline
% \end{tabular}
% \label{tab:running time}
% \end{table*}

We next discuss parameter selection (Sec.~\ref{ssec:parameter_selection}), computation efficiency (Sec.~\ref{ssec:efficiency}), and alternative designs (Sec.~\ref{ssec:alternatives}) for the topology density map.
We then discuss the limitations, and summarize the future work in the end (Sec.~\ref{ssec:limitation}).

\subsection{Parameter Selection}
\label{ssec:parameter_selection}

As discussed in Sec.~\ref{ssec:estimate_surface}, a kernel function is used as a decay factor when we estimate density on a 2D planar surface. 
% Equation~4 in Section~\ref{estimate_surface} describes this step.
% The function of this step has been mentioned in Section 5.2. which is:$$d(j) = Acc_t^P K(\frac{dis}{r})$$
% where $K$ represents the kernel function and $r$ is the bandwidth which is also known as search radius. 
% The idea of KDE is to assign different weights to different points. So in our method, it means that on the surface, the longer distance to the corresponding intersection the faster the decay.
Therefore, a reasonable decay function is particularly important for topology density map.
A reasonable decay effect can not only yield an intuitive visualization, but also help users understand the density distribution in urban areas and facilitate decision-making.
In contrast, a slow or fast decay parameter will increase the difficulty of observing the density distribution.

Two key parameters in KDE, \emph{i.e.}, kernel function $K$, and bandwidth $r$, affect the decay speed.
Many studies~\cite{bailey1995interactive,o2014geographic,o2007surface,schabenberger2017statistical} suggest that choosing a proper bandwidth is more important than a kernel function.
Therefore, we only compare two widely used kernel functions $-$ Gaussian and Sigmoid.
For each kernel function, we further compare five equally spaced bandwidths from 0.001 to 0.005.
\fzz{Bandwidth $r$ corresponds to the accessibility value as in Equation~\ref{equ:acc} measured between the target point on the 2D planar surface and the source POI.
We compute the density estimation when the accessibility value is within the range; otherwise, we set the density value to 0.}
% In our study, $r$~=~0.003 indicates that 320m approximately corresponds to the distance of the real-world map.}
We aim to find a good combination of kernel function and bandwidth for this study.

The comparison result is shown in Fig.~\ref{fig:parameter selection}.
Each sub-figure presents the density field of the region enclosed by the rectangle.
The vertices of the rectangle represent intersections, and edges represent road segments in a road network.
In each rectangle, two different POIs are positioned at the top-left (green POI) and bottom-right (purple POI) vertices. Here, we only compare differences of density distribution within the planar surface computed from different parameters, so the density distribution on road segments is not taken into consideration.
Fig.~\ref{fig:parameter selection} indicates that for the same POIs and road topology, different combinations of kernel function and bandwidth produce different visual effects.
In Fig.~\ref{fig:parameter selection}(a,b,f), the density decay is too fast, making it difficult to observe the density change, especially for points in the middle of each rectangle.
Consequently, it is difficult for users to perceive accessibility differences.
In Fig.~\ref{fig:parameter selection}(d,e,h,i,j), the density decay is too slow to distinguish the density values.
The results suggest to discard these parameter combinations. 
In addition, both Gaussian and Sigmoid kernels show similar trends of decaying too slow with small bandwidths, whilst too fast with big bandwidths.
The results reveal that different kernel functions have no significant impact on density estimation, but different bandwidths do.
This finding is consistent with those of previous studies~\cite{bailey1995interactive,o2014geographic,o2007surface,schabenberger2017statistical}.

On the basis of the results in Fig.~\ref{fig:parameter selection}, we eventually choose Gaussian as the kernel function and 0.003 as the bandwidth.
All experiments and case studies in this work are conducted using these parameters.
Nevertheless, the visual effects are also determined by many factors, including accessibility measurement, color perception, and study areas.
Different application scenarios may have different requirements for decay effect.
The purpose of this paper is to illustrate an accurate and intuitive method for density map generation.
Users can make their own choices in different cases.

\begin{table}[h]
\small
\caption{Running time (in milliseconds) comparison for topology density map and planar KDE by different number of POIs and image size.}
\centering
\begin{tabularx}{0.495\textwidth}{|
							>{\hsize=0.16\hsize}c|
							>{\hsize=0.16\hsize}X|
							>{\hsize=0.14\hsize}X|
							>{\hsize=0.16\hsize}X|
							>{\hsize=0.14\hsize}X|
							>{\hsize=0.16\hsize}X|
							>{\hsize=0.14\hsize}X|}
							
\hline
\multirow{2}{*}{Image Size} 
% \multirow{2}*{Screen Size}
% \multirow{2}*{\diagbox{Size of Image}{Applied Method}{Number of POIs}} 
        % \multirow{2}*{\diagbox {Image Size}{\textbf{\# POIs}}}
        & \multicolumn{2}{c|}{10 POIs}  & \multicolumn{2}{c|}{50 POIs} & \multicolumn{2}{c|}{100 POIs}\\ 
        \cline{2-7} 
 &{Topology} & {Planar} & {Topology} & {Planar} & {Topology} & {Planar} \\
 \hline	
{1920$\times$1200} & {2010.75} & {99.41}& {1992.04} & {144.21}& {2134.92}& {192.20}\\ \hline
{1280$\times$800} & {962.55} &{51.25} & {971.14} & {78.49}& {991.41} & {116.29} \\ \hline
{640$\times$400}  & {353.87} & {18.98}& {381.97}& {24.33} & {347.03}& {32.70}\\ \hline
\end{tabularx}
\vspace{-2mm}
\label{tab:running time}
\end{table}

\subsection{Computation Efficiency}
\label{ssec:efficiency}

\fzz{
NKDE only estimates the density along a 1D road network, and often the density fields are estimated using average travel speeds of road segments. 
In contrast, topology density map and planar KDE provide density estimation over an entire 2D planar surface.
Hence, NKDE would require much less computation time than topology density map and planar KDE.
For fairness, we only compare the computation efficiency of our approach with planar KDE and not with NKDE.
The experiment runs on a MacBook Pro 2.5 GHz Core i7 with a Radeon R9 M370X graphics card, and the density maps are rendered in a front-end web interface using Javascript.
To avoid interference of other factors, we adopt the same study area, traffic data, and POIs in both density maps.
Our topology density map takes pre-computed DAGs (Sec.~\ref{ssec:estimate_road}) starting from each POI to all intersections using Dijkstra shortest path algorithm as additional input.
% In addition, the process of computing the shortest path from each intersection to the corresponding POI is based on Dijkstra shortest path algorithm, so the computational complexity of this part is the same with the Dijkstra algorithm, which we strip this part and consider it as a part of data pre-processing from topology density map.

Table~\ref{tab:running time} shows the running times (in milliseconds) of the topology density map and planar KDE under different numbers of POIs and image sizes.
Overall, the table shows that the computation efficiency for planar KDE is affected by both image size and number of POIs, while that for the topology density map is mainly affected by image size.
This is consistent with our expectation, as planar KDE needs to measure density fields for 2D space (\emph{i.e.}, the image size) based on input POIs (see Equation~\ref{equ:planar_KDE}).
Instead, computation cost for the topology density map is mainly spent on extending density fields from road network to 2D space (see Equation~\ref{equ:density_final}), and the process is dependent on the number of intersections rather than number of POIs.
Moreover, with the same image size and number of POIs, planar KDE requires much less running time than the topology density map, especially when the number of POIs is small.
This is because there are in total 273 intersections in the study area, which is much more than that of POIs, and for each point in a region, we need to compute density fields extended from multiple neighboring intersections.

Note that the running time for the topology density map is measured scratch, which is much slower than that for planar KDE.
Nevertheless, in many real-world scenarios (\emph{e.g.}, case study 2), we can pre-compute a density estimation for an entire area based on existing POIs, then we only need to re-compute the density fields for a small region affected by a new POI.
This process can be finished in real-time since the number of intersections to be updated is small.
% In the case of the same POIs, the planar KDE has a slightly better performance than the topology density map, but in terms of the order of magnitude, the running time for planar KDE and topology density map belong to the same level.
% When comparing the influence of different sizes of images on the two methods, we find that for the same number of POIs, whether topology density map or planar KDE, the running time has a significant difference.
% This indicates that in comparison with planar KDE, both two methods (topology density map and planar KDE) will not increase the computational complexity due to the increase of POIs. The running time for both methods is related to the size of the rendered image, which is the larger the image, the longer the time. In this study, we do not focus on optimizing the process of rendering the image, but we may try more exploration to improve the efficiency of rendering in the future.
}

% \begin{table*}[]
% \begin{tabular}{|l|l|l|l|l|l|l|}
% \hline
%           & \multicolumn{4}{l|}{{ \textbf{5}}}                                                  & \multicolumn{4}{l|}{{ \textbf{10}}}& \multicolumn{4}{l|}{{ \textbf{20}}}                                                            \\ \hline
% \diagbox{Size of Image}{Number of POIs}          & { \textbf{5}}         & { \textbf{10}}      & { \textbf{20}}         & {\textbf{5}}          & { \textbf{10}}      & { \textbf{20}}        \\ \hline
% 1920*1200 & { 1403.12ms}     & { 1351.76ms} & { 1411.33ms} & { 1028.20ms} & { 1083.43ms}  & { 1161.22ms} \\ \hline
% 1280*800  & { 933.26ms} & {954.28ms}  & { 950.34ms}    & { 845.42ms}  & {787.21ms} & {802.68ms} \\ \hline
% 640*400   & {396.023ms}  & { 432.33ms} & { 478.29ms}  & { 361.67ms}  & { 384.54ms}  & { 364.62ms}   \\ \hline
% \end{tabular}
% \end{table*}

% \multirow{-2}{*}

\begin{figure}[h]
  \centering
  \includegraphics[width=\linewidth]{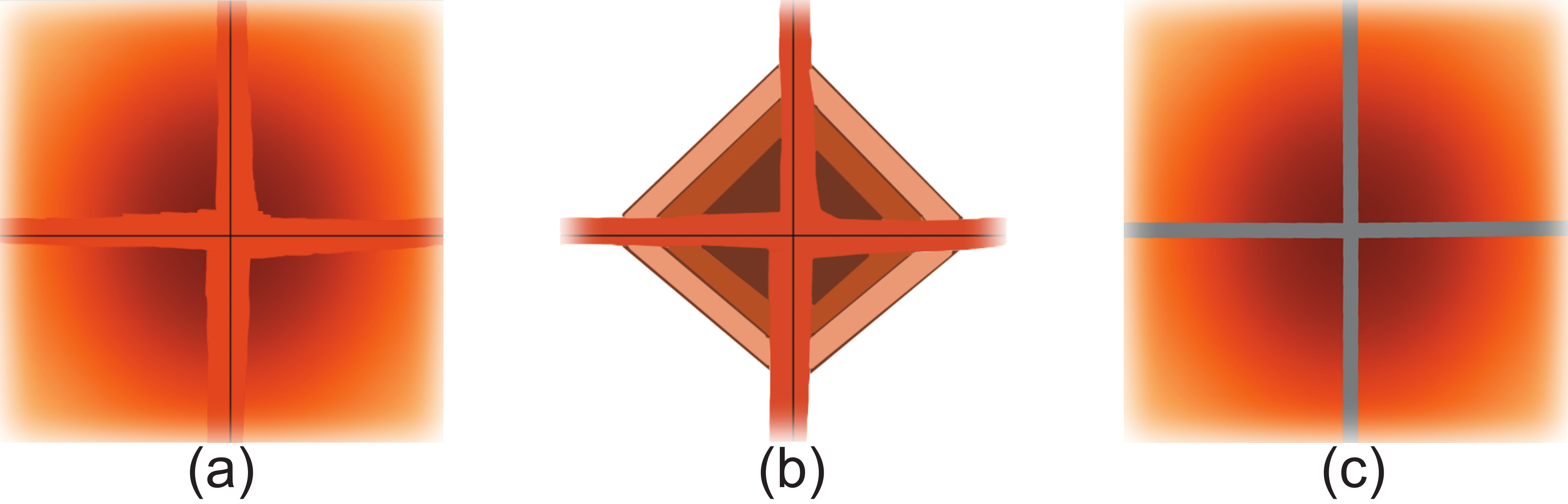}
  \vspace{-2em}
  \caption{Alternative designs to visualize density distribution: (a) topology density map used in this study, (b) a contour-based design, and (c) density map without showing density on roads.
  }~\label{fig: alternative design}
  \vspace{-2em}
%   \vspace{-0.2cm}
\end{figure}

\subsection{Alternative Design}
\label{ssec:alternatives}

Two alternative designs for visualizing scalar fields distributed in a 2D space are considered.
Fig.~\ref{fig: alternative design}(b) is our first attempt, which employs tapered lines with line thickness to visualize the density propagation along roads, and a contour map to visualize the density distribution in a 2D planar surface.
Each contour represents the region that can be accessed within a certain time.
However, after discussing the design choice with a domain expert (Expert C), she pointed out that a contour map can only depict discrete access times.
As in Fig.~\ref{fig: alternative design}(b), there are only three time intervals, and one can only retrieve which places can be accessed within these three different time intervals.
We can increase the number of time intervals, but the contour map will become chaotic when contours are dense. Consequently, finding useful information becomes difficult.
Moreover, visualizing the density fields is difficult for contour map when multiple POIs exist.
Fig.~\ref{fig: alternative design}(c) is another design that adopts density map to visualize density fields in a 2D planar surface. 
The density map is constructed in the same way as our topology density map.
However, the design adopts only straight lines to reflect road connections, rather than tapered lines to visualize density propagation along road segments.
Without tapered lines, density propagation along roads is difficult for users to follow.
% In this design, when computing the accessibility on the road we consider the road topology and traffic condition, but we do not visualize it on the road. This has similar result with our final version, but without visualizing the density on the road makes it difficult for users to understand the reason of density distribution.

Therefore, we choose topology density map coupled with tapered lines (Fig.~\ref{fig: alternative design}(a)) as our final design.

\subsection{Limitations and Future Work}
\label{ssec:limitation}
\emph{Limitations}. The case studies (Secs.~\ref{ssec:study1} \& \ref{ssec:study2}) and expert feedbacks (Sec.~\ref{ssec:interview}) confirm that topology density map satisfies the correctness and intuitiveness requirements (Sec.~\ref{ssec:limitation}). 
Nevertheless, improvement can still be made. 
% \textbf{Computation Efficiency.} 
First, our current implementation does not include the efficiency of generating the density map as the first priority.
We need to compute the density value for each pixel in the screen, which is rather time-consuming.
Several promising directions for improving the computing efficiency can be taken, such as GPU-based implementation of density computation~\cite{lampe_2011_interactive} and space partition using weighted Voronoi diagram~\cite{hoff1999fast}.
% In addition to algorithm optimization, we would like to make a trade-off between visual quality and efficiency in the future.
% \textbf{} 
Second, in the case studies, we assume that vehicles drive on a road segment at a constant speed, which is impossible in reality.
We make this simplified assumption because a more detailed traffic data is unavailable.
As suggested by domain experts, we would like to produce a more accurate density estimation with fine-grained speed information on roads. 
Third, due to the lack of building and walkway data in each zone, we only consider the direct connection between a location and its surrounding intersections when extending the density estimation from the road network to a planar surface, which again leads to inaccurate estimation.
For example, if a zone (\emph{e.g.}, a campus or housing estate) possesses only one entrance, the density estimation for a location in the zone shall include the location's distance to the entrance and the entrance's distance to road network.
We would like to fuse other kinds of urban data, such as land use and building types, to generate more accurate density estimation.
% If we can get detailed geographical and traffic data, then the distribution of density can be estimated more accurately from the entrance instead of intersections.

% \subsection{Future Work}
% \label{ssec:future}
% Previous studies on density map either focused on the estimation of Euclidean distance on the planar surface or the density estimation based on network distance on the road. However, our method not only covers the planar surface, but also takes the road topology and the traffic conditions into consideration.
\vspace{1mm}
\noindent
\emph{Future Work}.
In addition to improving the efficiency and accuracy of the topology density map construction, many fruitful directions for future work can be taken.
First, we would like to integrate topology density map into many real-world applications, \emph{e.g.}, a real-time risk monitoring system.
The system can provide interactive visualization for POI accessibility analysis under real-time traffic conditions.
We anticipate that such a system can assist policy makers in making appropriate decisions in case of emergency, such as to improve fire truck scheduling strategies in case of fire.
Second, providing temporal information in the context of a map view is a always challenging task in geographical information system.
Many studies adopt isochrone map, which however can only depict discrete temporal information; see the comparison of topology density map with contour map above.
For an example, Zeng et al.~\cite{zeng2014visualizing} used isochrone maps to depict accessible regions within 30 and 60 minutes using public transportation system.
However, detailed travel times can only be depicted with an additional isotime flow map view.
We plan to integrate temporal information on a map seamlessly using topology density map.
Last but not least, this work employs access time as the indicator to reflect POI accessibility.
In fact, topology density map can be applied in various scenarios as long as the analyses rely on certain topological structures, \emph{e.g.}, network connectivity.
We would like to exploit the potential of applying topology density map in those scenarios.
% For example, topology density map can be employed to study public transportation systems.
% Taking the topology of metro lines into consideration, domain experts can analyze the influence of building a new metro line or station on economy.
\section{Conclusion}
\label{sec:con}

In this paper, we have introduced \emph{Topology Density Map}, a new method for density estimation in the context of an urban environment.
Topology density map surpasses the performance of conventional planar KDE and NKDE in satisfying two specific requirements for urban data visualization and analysis:
First, the density fields estimation should be \emph{correct}, in order to support urban analysis and assist decision-making;
Second, the visualization should be \emph{intuitive} that depicts density fields over an entire 2D space instead of only 1D road network.
The correctness requirement is achieved by including complex road topology and dynamic traffic conditions into consideration, while the intuitiveness requirement is achieved by extending accessibility to arbitrary points in the 2D space.

To facilitate the construction process of topology density map, we first construct a series of DAGs for input POIs, of which a DAG propagates nonlinear accessibility from the source POI to all intersections in a road network.
Next, to extend density field estimation from road network to the entire planar surface, we identify key vertices in each DAG, compute and compare density field for every point assorted with each POI, and finally color code every point according to the computed density field and corresponding POI.
Topology density maps exhibit a weighted Voronoi diagram-like visual effect, which divides the 2D planar surface into different regions in accordance with POI accessibility.
Various parameters and alternative designs have been considered, and we choose an optimal setting that meets the intuitiveness requirement.
Two case studies of exploring hospital accessibility based on real-world traffic data in an urban area, together with positive feedbacks from three independent domain experts, approved the topology density map in terms of feasibility, usability, and effectiveness.
\section{acknowledgments}
\label{sec:ack}

We thank all the domain experts interviewed in this research. We also thank the reviewers for their comments and suggestions. The work is supported in part by the National Key Research and Development Plans (2019YFC0810705 and 2018YFC0807002) of P. R. China, National Natural Science Foundation of China (No. 61802388), HK RGC GRF grant 16208514 and HKUST SSC grant F0707.

\bibliographystyle{abbrv-doi}

\bibliography{template}
\end{document}